\documentclass[prd,aps,showpacs,epsf,floats]{revtex4}
\usepackage{amsfonts}
\usepackage{amsmath}

\input{tcilatex}
\begin{document}

\title{Repetition Versus Noiseless Quantum Codes For Correlated Errors}
\author{Carlo Cafaro}
\email{carlo.cafaro@unicam.it}
\affiliation{Dipartimento di Fisica, Universit\`{a} di Camerino, I-62032 Camerino, Italy}
\author{Stefano Mancini}
\email{stefano.mancini@unicam.it}
\affiliation{Dipartimento di Fisica, Universit\`{a} di Camerino, I-62032 Camerino, Italy}

\begin{abstract}
We study the performance of simple quantum error correcting codes with
respect to correlated noise errors characterized by a finite correlation
strength $\mu $. Specifically, we consider bit flip (phase flip) noisy
quantum memory channels and use repetition and noiseless quantum codes. We
characterize the performance of the codes by means of the entanglement
fidelity $\mathcal{F}\left( \mu \text{, }p\right) $ as function of the error
probability $p$ and degree of memory $\mu $. Finally, comparing the
entanglement fidelities of repetition and noiseless quantum codes, we find a
threshold $\mu ^{\ast }\left( p\right) $ for the correlation strength that
allows to select the code with better performance.
\end{abstract}

\maketitle

\section{Introduction}

Decoherence is the most important obstacle in quantum computing. It causes a
quantum computer to lose its quantum properties destroying its performance
advantages over a classical computer. Therefore, in order to maintain
quantum coherence in any computing system, it is important to remove the
unwanted\emph{\ }entanglement with its noisy environment. The unavoidable
interaction between the open quantum computing system and its environment
corrupts the information stored in the system and causes computational
errors that may lead to wrong outputs. In general, environments may be very
complex systems characterized by many uncontrollable degrees of freedom. The
principal task of quantum error correction (QEC, \cite{nielsen00, laflamme07}%
) is to tackle this decoherence problem. For a comprehensive introduction to
quantum error correction, we refer to the work presented in \cite{perimeter}%
. In summary, there exists two strategies to defend quantum coherence of a
processing against environmental noise. The first strategy is that of \emph{%
quantum error correcting codes} (QECC) \cite{knill97, calderbank97} where,
in analogy to classical information theory, quantum information is
stabilized by using redundant encoding and measurements. This is also known
as an active strategy. The second strategy is known as \emph{noiseless
quantum coding} ( also known as error avoiding quantum coding or decoherence
free subspaces (DFSs)) \cite{zanardi97, zanardi97+, lidar98, lidar99,
bacon00}. This is a passive strategy where quantum information is stabilized
by exploiting symmetry properties of the environment-induced noise for
suitable redundant encoding.

The formal mathematical description of the qubit-environment interaction is
usually given in terms of quantum channels. When noise errors act
independently on each qubit, we talk about memoryless (noisy quantum)
channels and independent error models. Instead, when noise errors do not
affect qubits independently but correlations between errors on different
qubits must be taken into consideration, we talk about memory (noisy
quantum) channels and correlated error models. Correlations between errors
may be considered either temporally over each use of a single channel, or
spatially between uses of many parallel channels. QECC were developed under
the assumption of i.i.d. (identically and independently distributed) errors.
Recent studies on the performance of quantum codes for memory channels
appear in \cite{arrigo08, shabani08}. In Ref. \cite{clemens04}, the
performance of some codes (CSS codes and\textbf{\ }$n$\textbf{-}qubit
repetition code) for spatially-correlated errors were studied and
characterized by means of the lowest order temporal expansion of the
fidelity of the density operator representing the quantum register after a
single application of error correction.

In this Letter, we study the performance of simple quantum error correcting
codes in the presence of correlated (classical-like) noise error models
characterized by a correlation strength. Specifically, we consider bit flip
(phase flip) noisy quantum memory channels and use repetition and noiseless
quantum codes. Although the error models considered are classical-like, we
can gain useful insights for extending error correction techniques to fully
quantum correlated error models. We characterize the performance of such
error correcting codes by means of the entanglement fidelity \cite%
{schumacher96} as function of the error probability and degree of memory.
Finally, comparing the entanglement fidelities of repetition codes and
noiseless quantum codes, we find a threshold for the correlation strength
that allows to select the code with better performance.

The layout of this Letter is as follows. In Section II, the algorithmic
structure of a basic quantum error correcting code and a brief description
of independent and correlated error models are presented. In Section III, we
introduce a simple error model in the presence of correlated errors.
Specifically, we consider bit flip (or phase flip) noisy quantum memory
channels. The performance of quantum error correcting codes is quantified by
means of the entanglement fidelity $\mathcal{F}_{RC}^{\left( n\right)
}\left( \mu \text{, }p\right) $ ($RC=$ repetition code; $n$ is the length of
the code) as function of the error probability $p$ and degree of memory $\mu 
$. In Section IV, we use odd and even-length error avoiding quantum error
correcting codes and characterize the performance of the codes by means of
the entanglement fidelity $\mathcal{F}_{DFS}^{\left( n\right) }\left( \mu 
\text{, }p\right) $ \cite{lidar03}). In Section V, we briefly discuss the
existence of threshold values for the correlation strength $\mu ^{\ast
}\left( p\right) $ that allows to select, for a fixed value of the dimension
of the coding space and the error probability $p$, the quantum error
correcting code with better performance and we present our final remarks.

\section{Quantum Error Correction and Error Models}

In this Section, we present the algorithmic structure of a basic quantum
error correcting code and briefly describe independent and correlated error
models in quantum computation.

\subsection{Algorithmic Structure of Quantum Error Correction}

\emph{Characterizing the Error Model}. We may deal with error models where
errors occur on single qubits independently (independent error models).
However, we may also consider models in which errors do not affect qubits
independently. In this case we have to take into account any correlation
between errors on different qubits (correlated error model). In any case,
the error model is completely described by the Kraus operators $\left\{
A_{j}\right\} $ defining the quantum error operation $\Lambda $. The
trace-preserving superoperator $\Lambda $ is described as follows,%
\begin{equation}
\rho \overset{\text{def}}{=}\left\vert \psi \right\rangle \left\langle \psi
\right\vert \overset{\Lambda }{\longrightarrow }\Lambda \left( \rho \right) 
\overset{\text{def}}{=}\sum_{j}A_{j}\left\vert \psi \right\rangle
\left\langle \psi \right\vert A_{j}^{\dagger }\text{.}  \label{eq}
\end{equation}

\emph{Introducing Redundancy by Choosing the Encoding}. The basic idea
behind the redundancy is that even when errors corrupt some of the qubits in
a codeword $\left\vert \psi _{\text{enc}}\right\rangle $ \cite{nielsen00},
the remaining qubits contain enough information so that the logical qubit $%
\left\vert \psi \right\rangle $, representing the quantum information to be
transmitted through the quantum noisy communication channel $\Lambda $, can
be recovered. In quantum error correction, encoding is implemented via a
unitary operator $U_{\text{enc}}$ that acts on the state we wish to encode,
tensored with an ancilla of some fixed number of qubits in some specified
initial state. The goal is to choose the encoding operation $U_{\text{enc}}$
in such a way that the behavior of these transformed errors allows us to
find a recovery operation $\mathcal{R}$ that gives back $\left\vert \psi
\right\rangle \left\langle \psi \right\vert \otimes \rho _{\text{noise}}$.
The encoding operation is described as follows,%
\begin{equation}
\left\vert \psi \right\rangle \overset{\text{tensor product}}{%
\longrightarrow }\left\vert \psi \right\rangle \otimes \left\vert 00\text{...%
}0\right\rangle \overset{\text{ }U_{\text{enc}}}{\longrightarrow }U_{\text{%
enc}}\left( \left\vert \psi \right\rangle \otimes \left\vert 00\text{...}%
0\right\rangle \right) \overset{\text{def}}{=}\left\vert \psi _{\text{enc}%
}\right\rangle \text{.}
\end{equation}%
In other words, we consider the tensor product between the logical qubit $%
\left\vert \psi \right\rangle $ and the ancilla qubit $\left\vert 00\text{...%
}0\right\rangle $, and then we encode $\left\vert \psi \right\rangle \otimes
\left\vert 00\text{...}0\right\rangle $ via a unitary operator $U_{\text{enc}%
}$, obtaining $U_{\text{enc}}\left( \left\vert \psi \right\rangle \otimes
\left\vert 00\text{...}0\right\rangle \right) \overset{\text{def}}{=}%
\left\vert \psi _{\text{enc}}\right\rangle $.

\emph{Finding a Procedure for Error Recovery}. The encoding operation $U_{%
\text{enc}}$ can be seen as a way of transforming the encoded errors $%
A_{j}^{\prime }$ so that their action on the codeword states $\left\vert
\psi _{\text{enc}}\right\rangle $ is recoverable. Before encoding, the
quantum error operation $\Lambda $ is defined as in (\ref{eq}. After
encoding, the new quantum error operation $A^{\prime }$ is defined as
follows,%
\begin{equation}
\rho _{\text{enc}}\overset{\text{def}}{=}\left\vert \psi _{\text{enc}%
}\right\rangle \left\langle \psi _{\text{enc}}\right\vert \overset{\Lambda
^{\prime }}{\longrightarrow }\Lambda ^{\prime }\left( \rho _{\text{enc}%
}\right) \overset{\text{def}}{=}\sum_{j}A_{j}^{\prime }\left\vert \left(
\psi _{\text{enc}}\right) \right\rangle \left\langle \psi _{\text{enc}%
}\right\vert A_{j}^{\prime \dagger }\text{.}
\end{equation}%
The noise operators $A_{j}^{\prime }$ act on the codeword $\left\vert \psi _{%
\text{enc}}\right\rangle $ which lives in a larger dimensional space than
that of the original quantum state $\left\vert \psi \right\rangle $. The
noise operators $A_{j}$ act on $\left\vert \psi \right\rangle $. Notice that
in general,%
\begin{equation}
\text{Tr}_{\text{anc}}\left[ \sum_{k}U_{\text{enc}}^{\dag }\left(
\sum_{j}A_{j}^{\prime }U_{\text{enc}}\left\vert \psi \right\rangle
\left\vert 00\text{...}0\right\rangle \left\langle 00\text{...}0\right\vert
\left\langle \psi \right\vert U_{\text{enc}}^{\dagger }A_{j}^{\prime \dagger
}\right) U_{\text{enc}}\right] \neq \left\vert \psi \right\rangle
\left\langle \psi \right\vert \text{.}
\end{equation}%
Therefore, if we encode the quantum information $\left\vert \psi
\right\rangle $, subject it to the noise $A_{j}^{\prime }$ and decode using
the inverse of the encoding operation, $U_{\text{enc}}^{\dagger }=$ $U_{%
\text{enc}}^{-1}\overset{\text{def}}{=}U_{\text{dec}}$, we will not always
recover the original state $\left\vert \psi \right\rangle $. To recover $%
\left\vert \psi \right\rangle $ we need to introduce an error recovery
operation $\mathcal{R}$ that has the effect of undoing enough of the noise $%
A_{j}^{\prime }$ on the codeword state $\left\vert \psi _{\text{enc}%
}\right\rangle $ so that after decoding and tracing out we are left with $%
\left\vert \psi \right\rangle $ \cite{laflamme07},%
\begin{equation}
\text{Tr}_{\text{anc}}\left[ \sum_{k}\mathcal{R}_{k}U_{\text{enc}}^{\dag
}\left( \sum_{j}A_{j}^{\prime }U_{\text{enc}}\left\vert \psi \right\rangle
\left\vert 00\text{...}0\right\rangle \left\langle 00\text{...}0\right\vert
\left\langle \psi \right\vert U_{\text{enc}}^{\dagger }A_{j}^{\prime \dagger
}\right) U_{\text{enc}}\mathcal{R}_{k}^{\dag }\right] =\left\vert \psi
\right\rangle \left\langle \psi \right\vert \text{.}  \label{1}
\end{equation}%
The design of a quantum error correcting code can be reduced to finding a
unitary encoding operator $U_{\text{enc}}$ and a recovery operation $%
\mathcal{R}$ so that, given an error model corresponding to a specified set
of error operators $A_{j}$, equation (\ref{1}) is valid. The action of the
recovery operator $\mathcal{R}$ may be interpreted as pushing all the noise
into the ancilla $\left\vert 00\text{...}0\right\rangle $ so that the errors
are eliminated when the ancilla is traced out.

\emph{Computing the Entanglement Fidelity}. In \cite{schumacher96},
Schumacher introduced the concept of entanglement fidelity as a useful
indicator of the efficiency of quantum error correcting codes. The
entanglement fidelity is defined for a mixed state $\rho =\sum_{i}p_{i}\rho
_{i}=$tr$_{\mathcal{H}_{R}}\left\vert \psi \right\rangle \left\langle \psi
\right\vert $ in terms of a purification $\left\vert \psi \right\rangle \in 
\mathcal{H}\otimes \mathcal{H}_{R}$ to a reference system $\mathcal{H}_{R}$.
The purification $\left\vert \psi \right\rangle $ encodes all of the
information in $\rho $. Entanglement fidelity is a measure of how well the
channel $\Lambda $ preserves the entanglement of the state $\mathcal{H}$
with its reference system $\mathcal{H}_{R}$. The entanglement fidelity is
defined as follows \cite{schumacher96},%
\begin{equation}
\mathcal{F}\left( \rho \text{, }\Lambda \right) \overset{\text{def}}{=}%
\left\langle \psi |\left( \Lambda \otimes I_{\mathcal{H}_{R}}\right) \left(
\left\vert \psi \right\rangle \left\langle \psi \right\vert \right) |\psi
\right\rangle \text{,}
\end{equation}%
where $\left\vert \psi \right\rangle $ is any purification of $\rho $, $I_{%
\mathcal{H}_{R}}$ is the identity map on $\mathcal{M}\left( \mathcal{H}%
_{R}\right) $ and $\Lambda \otimes I_{\mathcal{H}_{R}}$ is the evolution
operator extended to the space $\mathcal{H}\otimes \mathcal{H}_{R}$, space
on which $\rho $ has been purified. If the quantum operation $\Lambda $ is
written in terms of its Kraus operator elements $\left\{ A_{k}\right\} $ as, 
$\Lambda \left( \rho \right) =\sum_{k}A_{k}\rho A_{k}^{\dagger }$, then it
can be shown that \cite{nielsen96}, 
\begin{equation}
\mathcal{F}\left( \rho \text{, }\Lambda \right) =\sum_{k}\text{tr}\left(
A_{k}\rho \right) \text{tr}\left( A_{k}^{\dagger }\rho \right)
=\sum_{k}\left\vert \text{tr}\left( \rho A_{k}\right) \right\vert ^{2}\text{.%
}
\end{equation}%
This expression for the entanglement fidelity is very useful for explicit
calculations. Finally, assuming that%
\begin{equation}
\Lambda :\mathcal{M}\left( \mathcal{H}\right) \ni \rho \longmapsto \Lambda
\left( \rho \right) =\sum_{k}A_{k}\rho A_{k}^{\dagger }\in \mathcal{M}\left( 
\mathcal{H}\right) \text{, dim}_{%
\mathbb{C}
}\mathcal{H=}N  \label{pla1}
\end{equation}%
and choosing a purification described by a maximally entangled unit vector $%
\left\vert \psi \right\rangle \in \mathcal{H}\otimes \mathcal{H}$ for the
mixed state $\rho =\frac{1}{\text{dim}_{%
\mathbb{C}
}\mathcal{H}}I_{\mathcal{H}}$ , we obtain%
\begin{equation}
\mathcal{F}\left( \frac{1}{N}I_{\mathcal{H}}\text{, }\Lambda \right) =\frac{1%
}{N^{2}}\sum_{k}\left\vert \text{tr}A_{k}\right\vert ^{2}\text{.}
\label{nfi}
\end{equation}%
The expression\ in (\ref{nfi}) represents the entanglement fidelity when no
error correction is performed on the noisy channel $\Lambda $ in (\ref{pla1}%
). In this\ Letter, we will follow the general theoretical framework
describing requirements for quantum error correcting codes presented in \cite%
{knill97}.

\subsection{Error Models}

To introduce noise models, we assume a quantum channel (CPTP map) $\Lambda $
on $n$-uses to be expressible by means of the following Kraus decomposition,%
\begin{equation}
\Lambda ^{\left( n\right) }\left( \rho \right) =\sum_{i_{1}\text{,..., }%
i_{n}}p\left( i_{n}\text{, }i_{n-1}\text{,..., }i_{1}\right) \left(
A_{i_{n}}\otimes \text{...}\otimes A_{i_{1}}\right) \rho \left(
A_{i_{n}}\otimes \text{...}\otimes A_{i_{1}}\right) ^{\dagger }\text{.}
\end{equation}%
If the probability $p\left( i_{n}\text{, }i_{n-1}\text{,..., }i_{1}\right) $
is factorized in the product of $n$-independent probabilities, $p\left( i_{n}%
\text{, }i_{n-1}\text{,..., }i_{1}\right) =\dprod\limits_{l=1}^{n}p\left(
i_{l}\right) $, we are in the presence of a memoryless channel and $\Lambda
^{\left( n\right) }\left( \rho \right) =\Lambda ^{\otimes n}\left( \rho
\right) \overset{\text{def}}{=}\Lambda \otimes $...$\otimes \Lambda $. On
the contrary if $p\left( i_{n}\text{, }i_{n-1}\text{,..., }i_{1}\right) $ is
not separable in the product of of $n$-independent probabilities, then $%
\Lambda $ is a memory channel with $\Lambda ^{\left( n\right) }\left( \rho
\right) \neq \Lambda ^{\otimes n}\left( \rho \right) $. For instance, a very
important class of quantum memory channels is described by the Markovian
correlated noise channels of length $n$,%
\begin{equation}
\Lambda ^{\left( n\right) }\left( \rho \right) \overset{\text{def}}{=}%
\sum_{i_{1}\text{,..., }i_{n}}p\left( i_{n}\text{\TEXTsymbol{\vert}}%
i_{n-1}\right) p\left( i_{n-1}\text{\TEXTsymbol{\vert}}i_{n-2}\right) \text{%
...}p\left( i_{2}\text{\TEXTsymbol{\vert}}i_{1}\right) p_{i_{1}}\left(
A_{i_{n}}\otimes \text{...}\otimes A_{i_{1}}\right) \rho \left(
A_{i_{n}}\otimes \text{...}\otimes A_{i_{1}}\right) ^{\dagger }\text{,}
\end{equation}%
with $p\left( i_{l}\text{\TEXTsymbol{\vert}}i_{l-1}\right) \overset{\text{def%
}}{=}\left( 1-\mu \right) p_{i_{l}}+\mu \delta _{i_{l}\text{, }i_{l-1}}$, $%
\forall l=1$,..., $n$. The correlation parameter $\mu $ describes the degree
of memory of the channel considered.

\section{Repetition Codes for Correlated Bit Flip}

In this Section, we introduce a simple error model in the presence of
correlated errors. Specifically, we consider bit flip (or phase flip) noisy
quantum memory channels and QEC is performed via odd and even repetition
codes (RC) \cite{knill02}. Although the error models considered are
classical in nature, from this preliminary work we hope to gain useful
insights for extending error correction techniques to quantum correlated
error models. The performance of quantum error correcting codes is
quantified by means of the entanglement fidelity $\mathcal{F}_{RC}^{\left(
n\right) }\left( \mu \text{, }p\right) $ as function of the error
probability $p$ and degree of memory $\mu $.

\subsection{CASE, $n_{\text{odd}}=3$}

Consider $n$ qubits and correlated errors in a bit flip quantum channel,%
\begin{equation}
\Lambda ^{(n)}(\rho )\overset{\text{def}}{=}\sum_{i_{1}\text{,..., }%
i_{n}=0}^{1}p_{i_{n}|i_{n-1}}p_{i_{n-1}|i_{n-2}}\text{...}%
p_{i_{2|i_{1}}}p_{i_{1}}\left( A_{i_{n}}\otimes \text{...}\otimes
A_{i_{1}}\right) \rho \left( A_{i_{n}}\otimes \text{...}\otimes
A_{i_{1}}\right) ^{\dagger }\text{,}
\end{equation}%
where $A_{0}\overset{\text{def}}{=}I$, $A_{1}\overset{\text{def}}{=}X$ are
Pauli operators. Furthermore,%
\begin{equation}
p_{i_{k}|i_{j}}=(1-\mu )p_{i_{k}}+\mu \delta _{i_{k}\text{, }i_{j}}\text{,}%
\quad p_{i_{k}=0}=1-p\text{,}\;p_{i_{k}=1}=p\text{,}  \label{aaa1}
\end{equation}%
with,%
\begin{equation}
\sum_{i_{1}\text{,..., }i_{n}=0}^{1}p_{i_{n}|i_{n-1}}p_{i_{n-1}|i_{n-2}}%
\text{...}p_{i_{2|i_{1}}}p_{i_{1}}=1\text{.}
\end{equation}%
To simplify our notation, we may assume that $A_{i_{n}}\otimes $...$\otimes
A_{i_{1}}\equiv $ $A_{i_{n}}$...$A_{i_{1}}$.

\emph{Error Operators}. In the simplest example, consider three qubits ($n=3$%
) and correlated errors in a bit flip channel,%
\begin{equation}
\Lambda ^{(3)}(\rho )\overset{\text{def}}{=}\sum_{i_{1}\text{, }i_{2}\text{, 
}i_{3}=0}^{1}p_{i_{3}|i_{2}}p_{i_{2}|i_{1}}p_{i_{1}}\left[
A_{i_{3}}A_{i_{2}}A_{i_{1}}\rho A_{i_{1}}^{\dag }A_{i_{2}}^{\dag
}A_{i_{3}}^{\dag }\right] \text{, with}\sum_{i_{1}\text{, }i_{2}\text{, }%
i_{3}=0}^{1}p_{i_{3}|i_{2}}p_{i_{2}|i_{1}}p_{i_{1}}=1\text{.}  \label{bit}
\end{equation}%
Substituting (\ref{aaa1}) in (\ref{bit}), it follows that the error
superoperator $\mathcal{A}$ associated to channel (\ref{bit}) is defined in
terms of the following error operators,%
\begin{equation}
\mathcal{A}\longleftrightarrow \left\{ A_{0}^{\prime }\text{,.., }%
A_{7}^{\prime }\right\} \text{ with }\Lambda ^{(3)}(\rho )\overset{\text{def}%
}{=}\dsum\limits_{k=0}^{7}A_{k}^{\prime }\rho A_{k}^{\prime \dagger }\text{
and, }\dsum\limits_{k=0}^{7}A_{k}^{\prime \dagger }A_{k}^{\prime
}=I_{8\times 8}\text{.}  \label{nota}
\end{equation}%
In an explicit way, the error operators $\left\{ A_{0}^{\prime }\text{,.., }%
A_{7}^{\prime }\right\} $ are given by,%
\begin{eqnarray}
A_{0}^{\prime } &=&\sqrt{\tilde{p}_{0}^{\left( 3\right) }}I^{1}\otimes
I^{2}\otimes I^{3}\text{, }A_{1}^{\prime }=\sqrt{\tilde{p}_{1}^{\left(
3\right) }}X^{1}\otimes I^{2}\otimes I^{3}\text{, }A_{2}^{\prime }=\sqrt{%
\tilde{p}_{2}^{\left( 3\right) }}I^{1}\otimes X^{2}\otimes I^{3}\text{, } 
\notag \\
&&  \notag \\
A_{3}^{\prime } &=&\sqrt{\tilde{p}_{3}^{\left( 3\right) }}I^{1}\otimes
I^{2}\otimes X^{3}\text{, }A_{4}^{\prime }=\sqrt{\tilde{p}_{4}^{\left(
3\right) }}X^{1}\otimes X^{2}\otimes I^{3}\text{, }A_{5}^{\prime }=\sqrt{%
\tilde{p}_{5}^{\left( 3\right) }}X^{1}\otimes I^{2}\otimes X^{3}\text{,} 
\notag \\
&&  \notag \\
\text{ }A_{6}^{\prime } &=&\sqrt{\tilde{p}_{6}^{\left( 3\right) }}%
I^{1}\otimes X^{2}\otimes X^{3}\text{, }A_{7}^{\prime }=\sqrt{\tilde{p}%
_{7}^{\left( 3\right) }}X^{1}\otimes X^{2}\otimes X^{3}\text{,}
\end{eqnarray}%
where the coefficients $\tilde{p}_{k}^{\left( 3\right) }$ for $k=1$,.., $7$
are given by,%
\begin{eqnarray}
\tilde{p}_{0}^{\left( 3\right) } &=&p_{00}^{2}p_{0}\text{, }\tilde{p}%
_{1}^{\left( 3\right) }=p_{00}p_{10}p_{0}\text{, }\tilde{p}_{2}^{\left(
3\right) }=p_{01}p_{10}p_{0}\text{, }\tilde{p}_{3}^{\left( 3\right)
}=p_{00}p_{01}p_{1}\text{,}  \notag \\
&&  \notag \\
\text{ }\tilde{p}_{4}^{\left( 3\right) } &=&p_{10}p_{11}p_{0}\text{, }\tilde{%
p}_{5}^{\left( 3\right) }=p_{01}p_{10}p_{1}\text{, }\tilde{p}_{6}^{\left(
3\right) }=p_{01}p_{11}p_{1}\text{, }\tilde{p}_{7}^{\left( 3\right)
}=p_{11}^{2}p_{1}\text{, }  \label{usa1}
\end{eqnarray}%
with,%
\begin{eqnarray}
p_{0} &=&\left( 1-p\right) \text{, }p_{1}=p\text{, }p_{00}=\left( \left(
1-\mu \right) \left( 1-p\right) +\mu \right) \text{, }  \notag \\
&&  \notag \\
p_{01} &=&\left( 1-\mu \right) \left( 1-p\right) \text{, }p_{10}=\left(
1-\mu \right) p\text{, }p_{11}=\left( \left( 1-\mu \right) p+\mu \right) 
\text{.}  \label{usa2}
\end{eqnarray}

\emph{Encoding and Decoding Operators}. Consider a repetition code that
encodes $1$ logical qubit into $3$-physical qubits. We have,%
\begin{equation}
\left\vert 0\right\rangle \overset{\text{Tensoring}}{\longrightarrow }%
\left\vert 0\right\rangle \otimes \left\vert 00\right\rangle =\left\vert
000\right\rangle \overset{\text{def}}{=}\left\vert 0_{\text{L}}\right\rangle 
\text{, }\left\vert 1\right\rangle \overset{\text{tensoring}}{%
\longrightarrow }\left\vert 1\right\rangle \otimes \left\vert
00\right\rangle =\left\vert 100\right\rangle \overset{U_{\text{CNOT}%
}^{12}\otimes I^{3}}{\longrightarrow }\left\vert 110\right\rangle \overset{%
U_{\text{CNOT}}^{13}\otimes I^{2}}{\longrightarrow }\left\vert
111\right\rangle \overset{\text{def}}{=}\left\vert 1_{\text{L}}\right\rangle 
\text{.}  \label{placs}
\end{equation}%
The operator $U_{\text{CNOT}}^{ij}$ is the CNOT\ gate from qubit $i$ to $j$
defined as,%
\begin{equation}
U_{\text{CNOT}}^{ij}\overset{\text{def}}{=}\frac{1}{2}\left[ \left(
I^{i}+Z^{i}\right) \otimes I^{j}+\left( I^{i}-Z^{i}\right) \otimes X^{j}%
\right] \text{.}  \label{you}
\end{equation}%
Finally, the encoding operator $U_{\text{enc}}$ such that $U_{\text{enc}%
}\left\vert 000\right\rangle =\left\vert 000\right\rangle $ and $U_{\text{enc%
}}\left\vert 100\right\rangle =\left\vert 111\right\rangle $ is defined as,%
\begin{equation}
U_{\text{enc}}\overset{\text{def}}{=}\left( U_{\text{CNOT}}^{13}\otimes
I^{2}\right) \circ \left( U_{\text{CNOT}}^{12}\otimes I^{3}\right) \text{.}
\label{bitencoding}
\end{equation}

\emph{Recovery Operators}. The set of error operators satisfying the
detectability condition \cite{knill02}, $P_{\mathcal{C}}A_{k}^{\prime }P_{%
\mathcal{C}}=\lambda _{A_{k}^{\prime }}P_{\mathcal{C}}$, where $P_{\mathcal{C%
}}=\left\vert 0_{L}\right\rangle \left\langle 0_{L}\right\vert +$ $%
\left\vert 1_{L}\right\rangle \left\langle 1_{L}\right\vert $ is the
projector operator on the code subspace $\mathcal{C}=Span\left\{ \left\vert
0_{L}\right\rangle \text{, }\left\vert 1_{L}\right\rangle \right\} $ is
given by,%
\begin{equation}
\mathcal{A}_{\text{detectable}}=\left\{ A_{0}^{\prime }\text{, }%
A_{1}^{\prime }\text{, }A_{2}^{\prime }\text{, }A_{3}^{\prime }\text{, }%
A_{4}^{\prime }\text{, }A_{5}^{\prime }\text{, }A_{6}^{\prime }\right\}
\subseteq \mathcal{A}\text{.}
\end{equation}%
The only non-detectable error is $A_{7}^{\prime }$. Furthermore, since all
the detectable errors are invertible, the set of correctable errors is such
that $\mathcal{A}_{\text{correctable}}^{\dagger }\mathcal{A}_{\text{%
correctable}}$ is detectable. It follows then that,%
\begin{equation}
\mathcal{A}_{\text{correctable}}=\left\{ A_{0}^{\prime }\text{, }%
A_{1}^{\prime }\text{, }A_{2}^{\prime }\text{, }A_{3}^{\prime }\right\}
\subseteq \mathcal{A}_{\text{detectable}}\subseteq \mathcal{A}\text{.}
\end{equation}%
The action of the correctable error operators $\mathcal{A}_{\text{correctable%
}}$ on the codewords $\left\vert 0_{L}\right\rangle $ and $\left\vert
1_{L}\right\rangle $ is given by,%
\begin{eqnarray}
\left\vert 0_{L}\right\rangle &\rightarrow &A_{0}^{\prime }\left\vert
0_{L}\right\rangle =\sqrt{\tilde{p}_{0}^{\left( 3\right) }}\left\vert
000\right\rangle \text{, }A_{1}^{\prime }\left\vert 0_{L}\right\rangle =%
\sqrt{\tilde{p}_{1}^{\left( 3\right) }}\left\vert 100\right\rangle \text{, }%
A_{2}^{\prime }\left\vert 0_{L}\right\rangle =\sqrt{\tilde{p}_{2}^{\left(
3\right) }}\left\vert 010\right\rangle \text{, }A_{3}^{\prime }\left\vert
0_{L}\right\rangle =\sqrt{\tilde{p}_{3}^{\left( 3\right) }}\left\vert
001\right\rangle \text{ }  \notag \\
&&  \notag \\
\left\vert 1_{L}\right\rangle &\rightarrow &A_{0}^{\prime }\left\vert
1_{L}\right\rangle =\sqrt{\tilde{p}_{0}^{\left( 3\right) }}\left\vert
111\right\rangle \text{, }A_{1}^{\prime }\left\vert 1_{L}\right\rangle =%
\sqrt{\tilde{p}_{1}^{\left( 3\right) }}\left\vert 011\right\rangle \text{, }%
A_{2}^{\prime }\left\vert 1_{L}\right\rangle =\sqrt{\tilde{p}_{2}^{\left(
3\right) }}\left\vert 101\right\rangle \text{, }A_{3}^{\prime }\left\vert
1_{L}\right\rangle =\sqrt{\tilde{p}_{3}^{\left( 3\right) }}\left\vert
110\right\rangle \text{.}  \label{ea}
\end{eqnarray}%
The two \ four-dimensional orthogonal subspaces $\mathcal{V}^{0_{L}}$ and $%
\mathcal{V}^{1_{L}}$ of $\mathcal{H}_{2}^{3}$ generated by the action of $%
\mathcal{A}_{\text{correctable}}$ on $\left\vert 0_{L}\right\rangle $ and $%
\left\vert 1_{L}\right\rangle $ are given by,%
\begin{equation}
\mathcal{V}^{0_{L}}=Span\left\{ \left\vert v_{1}^{0_{L}}\right\rangle
=\left\vert 000\right\rangle \text{,}\left\vert v_{2}^{0_{L}}\right\rangle
=\left\vert 100\right\rangle \text{, }\left\vert v_{3}^{0_{L}}\right\rangle
=\left\vert 010\right\rangle \text{, }\left\vert v_{4}^{0_{L}}\right\rangle
=\left\vert 001\right\rangle \text{ }\right\} \text{,}  \label{span1}
\end{equation}%
and,%
\begin{equation}
\mathcal{V}^{1_{L}}=Span\left\{ \left\vert v_{1}^{1_{L}}\right\rangle
=\left\vert 111\right\rangle \text{,}\left\vert v_{2}^{1_{L}}\right\rangle
=\left\vert 011\right\rangle \text{, }\left\vert v_{3}^{1_{L}}\right\rangle
=\left\vert 101\right\rangle \text{, }\left\vert v_{4}^{1_{L}}\right\rangle
=\left\vert 110\right\rangle \right\} \text{,}  \label{span2}
\end{equation}%
respectively. Notice that $\mathcal{V}^{0_{L}}\oplus \mathcal{V}^{1_{L}}=%
\mathcal{H}_{2}^{3}$. The recovery superoperator $\mathcal{R}\leftrightarrow
\left\{ R_{l}\right\} $ with $l=1$,..,$4$ is defined as \cite{knill97},%
\begin{equation}
R_{l}\overset{\text{def}}{=}V_{l}\sum_{i=0}^{1}\left\vert
v_{l}^{i_{L}}\right\rangle \left\langle v_{l}^{i_{L}}\right\vert \text{,}
\label{recovery}
\end{equation}%
where the unitary operator $V_{l}$ is such that $V_{l}\left\vert
v_{l}^{i_{L}}\right\rangle =\left\vert i_{L}\right\rangle $ for $i\in
\left\{ 0\text{, }1\right\} $. Substituting (\ref{span1}) and (\ref{span2})
into (\ref{recovery}), it follows that the four recovery operators $\left\{
R_{1}\text{, }R_{2}\text{, }R_{3}\text{, }R_{4}\right\} $ are given by,%
\begin{eqnarray}
R_{1} &=&\left\vert 0_{L}\right\rangle \left\langle 0_{L}\right\vert
+\left\vert 1_{L}\right\rangle \left\langle 1_{L}\right\vert \text{, }%
R_{2}=\left\vert 0_{L}\right\rangle \left\langle 100\right\vert +\left\vert
1_{L}\right\rangle \left\langle 011\right\vert \text{,}  \notag \\
&&  \notag \\
\text{ }R_{3} &=&\left\vert 0_{L}\right\rangle \left\langle 010\right\vert
+\left\vert 1_{L}\right\rangle \left\langle 101\right\vert \text{, }%
R_{4}=\left\vert 0_{L}\right\rangle \left\langle 001\right\vert +\left\vert
1_{L}\right\rangle \left\langle 110\right\vert \text{.}  \label{rec}
\end{eqnarray}%
Using simple algebra, it turns out that the $8\times 8$ matrix
representation $\left[ R_{l}\right] $ with $l=1$,..,$4$ of the recovery
operators is given by,%
\begin{equation}
\left[ R_{1}\right] =E_{11}+E_{88}\text{, }\left[ R_{2}\right] =E_{12}+E_{87}%
\text{, }\left[ R_{3}\right] =E_{13}+E_{86}\text{, }\left[ R_{4}\right]
=E_{14}+E_{85}\text{, }
\end{equation}%
where $E_{ij}$ is the $8\times 8$ matrix where the only non-vanishing
element is the one located in the $ij$-position and it equals $1$. It
follows that $\mathcal{R}\leftrightarrow \left\{ R_{l}\right\} $ is indeed a
trace preserving quantum operation since,%
\begin{equation}
\sum_{l=1}^{4}R_{l}^{\dagger }R_{l}=I_{8\times 8}\text{.}
\end{equation}%
Considering this recovery operation $\mathcal{R}$, the map $\Lambda ^{\left(
3\right) }\left( \rho \right) $ in (\ref{nota}) becomes,%
\begin{equation}
\Lambda _{\text{recover}}^{\left( 3\right) }\left( \rho \right) \equiv
\left( \mathcal{R\circ }\Lambda ^{(3)}\right) \left( \rho \right) \overset{%
\text{def}}{=}\sum_{k=0}^{7}\dsum\limits_{l=1}^{4}\left( R_{l}A_{k}^{\prime
}\right) \rho \left( R_{l}A_{k}^{\prime }\right) ^{\dagger }\text{.}
\label{pla2}
\end{equation}

\emph{Entanglement Fidelity}. We want to describe the action of $\mathcal{%
R\circ }\Lambda ^{(3)}$ restricted to the code subspace $\mathcal{C}$.
Therefore, we compute the $2\times 2$ matrix representation $\left[
R_{l}A_{k}^{\prime }\right] _{|\mathcal{C}}$ of each $R_{l}A_{k}^{\prime }$
with $l=1$,.., $4$ and $k=0$,.., $7$ where,%
\begin{equation}
\left[ R_{l}A_{k}^{\prime }\right] _{|\mathcal{C}}\overset{\text{def}}{=}%
\left( 
\begin{array}{cc}
\left\langle 0_{L}|R_{l}A_{k}^{\prime }|0_{L}\right\rangle & \left\langle
0_{L}|R_{l}A_{k}^{\prime }|1_{L}\right\rangle \\ 
\left\langle 1_{L}|R_{l}A_{k}^{\prime }|0_{L}\right\rangle & \left\langle
1_{L}|R_{l}A_{k}^{\prime }|1_{L}\right\rangle%
\end{array}%
\right) \text{.}  \label{sopra1}
\end{equation}%
Substituting (\ref{ea}) and (\ref{rec}) into (\ref{sopra1}), it turns out
that the only matrices $\left[ R_{l}A_{k}^{\prime }\right] _{|\mathcal{C}}$
with non-vanishing trace are given by,%
\begin{eqnarray}
\left[ R_{1}A_{0}^{\prime }\right] _{|\mathcal{C}} &=&\sqrt{\tilde{p}%
_{0}^{\left( 3\right) }}\left( 
\begin{array}{cc}
1 & 0 \\ 
0 & 1%
\end{array}%
\right) \text{, }\left[ R_{2}A_{1}^{\prime }\right] _{|\mathcal{C}}=\sqrt{%
\tilde{p}_{1}^{\left( 3\right) }}\left( 
\begin{array}{cc}
1 & 0 \\ 
0 & 1%
\end{array}%
\right) \text{,}  \notag \\
&&  \notag \\
\text{ }\left[ R_{3}A_{2}^{\prime }\right] _{|\mathcal{C}} &=&\sqrt{\tilde{p}%
_{2}^{\left( 3\right) }}\left( 
\begin{array}{cc}
1 & 0 \\ 
0 & 1%
\end{array}%
\right) \text{, }\left[ R_{4}A_{3}^{\prime }\right] _{|\mathcal{C}}=\sqrt{%
\tilde{p}_{3}^{\left( 3\right) }}\left( 
\begin{array}{cc}
1 & 0 \\ 
0 & 1%
\end{array}%
\right) \text{.}
\end{eqnarray}%
Therefore, the entanglement fidelity $\mathcal{F}_{RC}^{\left( 3\right)
}\left( \mu \text{, }p\right) $ defined as,%
\begin{equation}
\mathcal{F}_{RC}^{\left( 3\right) }\left( \mu \text{, }p\right) \overset{%
\text{def}}{=}\mathcal{F}^{\left( 3\right) }\left( \frac{1}{2}I_{2\times 2}%
\text{, }\mathcal{R\circ }\Lambda ^{(3)}\right) =\frac{1}{\left( 2\right)
^{2}}\sum_{k=0}^{7}\dsum\limits_{l=1}^{4}\left\vert \text{tr}\left( \left[
R_{l}A_{k}^{\prime }\right] _{|\mathcal{C}}\right) \right\vert ^{2}\text{,}
\label{fidel}
\end{equation}%
results,%
\begin{equation}
\mathcal{F}_{RC}^{\left( 3\right) }\left( \mu \text{, }p\right) =\tilde{p}%
_{0}^{\left( 3\right) }+\tilde{p}_{1}^{\left( 3\right) }+\tilde{p}%
_{2}^{\left( 3\right) }+\tilde{p}_{3}^{\left( 3\right) }\text{.}
\label{usa3}
\end{equation}%
The expression for $\mathcal{F}_{RC}^{\left( 3\right) }\left( \mu \text{, }%
p\right) $ in (\ref{fidel}) represents the entanglement fidelity quantifying
the performance of the error correction scheme provided by the repetition
code here considered. The quantum operation $\mathcal{R\circ }\Lambda ^{(3)}$
appearing in (\ref{fidel}) is defined in equation (\ref{pla2}) and the
recovery operators $R_{l}$ are explicitly given in (\ref{rec}). The action of%
\textbf{\ }$R_{l}A_{k}^{\prime }$ in (\ref{fidel}) is restricted to the code
space\textbf{\ }$\mathcal{C}$ defined in (\ref{placs}).

Substituting (\ref{usa1}) and (\ref{usa2}) into (\ref{usa3}), we finally
obtain%
\begin{equation}
\mathcal{F}_{RC}^{\left( 3\right) }\left( \mu \text{, }p\right) =\mu
^{2}\left( 2p^{3}-3p^{2}+p\right) +\mu \left( -4p^{3}+6p^{2}-2p\right)
+\left( 2p^{3}-3p^{2}+1\right) \text{.}
\end{equation}%
Notice that for a vanishing degree of memory $\mu $, the entanglement
fidelity becomes,%
\begin{equation}
\mathcal{F}_{RC}^{\left( 3\right) }\left( 0\text{, }p\right) =\sum_{m=0}^{1}%
\binom{3}{m}p^{m}\left( 1-p\right) ^{3-m}=2p^{3}-3p^{2}+1\text{.}
\end{equation}

\emph{Remarks on the coding for phase flip memory channels}. The code for
the phase flip channel has the same characteristics as the code for the bit
flip channel. These two channels are unitarily equivalent since there is a
unitary operator, the Hadamard gate $H$, such that the action of one channel
is the same as the other, provided the first channel is preceded by $H$ and
followed by $H^{\dagger }$ \cite{nielsen00},%
\begin{equation}
\Lambda ^{\text{phase}}\left( \rho \right) \overset{\text{def}}{=}\left(
H\circ \Lambda ^{\text{bit}}\circ H^{\dagger }\right) \left( \rho \right)
=\left( 1-p\right) \rho +pZ\rho Z\text{,}
\end{equation}%
where,%
\begin{equation}
\Lambda ^{\text{bit}}\left( \rho \right) \overset{\text{def}}{=}\left(
1-p\right) \rho +pX\rho X\text{ and, }H\overset{\text{def}}{=}\frac{1}{\sqrt{%
2}}\left( 
\begin{array}{cc}
1 & 1 \\ 
1 & -1%
\end{array}%
\right) \text{.}
\end{equation}%
These operations may be trivially incorporated into the encoding and
error-correction operations. The encoding for the phase flip channel is
performed in two steps: i) first, we encode in three qubits exactly as for
the bit flip channel; ii) second, we apply a Hadamard gate to each qubit,%
\begin{equation}
\left\vert 0\right\rangle \overset{\text{tensor}}{\longrightarrow }%
\left\vert 000\right\rangle \overset{U_{\text{enc}}^{\text{bit}}}{%
\longrightarrow }\left\vert 000\right\rangle \overset{H^{\otimes 3}}{%
\longrightarrow }\left\vert 0_{\text{L}}\right\rangle \overset{\text{def}}{=}%
\left\vert +++\right\rangle \text{, }\left\vert 1\right\rangle \overset{%
\text{tensor}}{\longrightarrow }\left\vert 100\right\rangle \overset{U_{%
\text{enc}}^{\text{bit}}}{\longrightarrow }\left\vert 111\right\rangle 
\overset{H^{\otimes 3}}{\longrightarrow }\left\vert 1_{\text{L}%
}\right\rangle \overset{\text{def}}{=}\left\vert ---\right\rangle \text{,}
\end{equation}%
where,%
\begin{equation}
\left( 
\begin{array}{c}
\left\vert +\right\rangle \\ 
\left\vert -\right\rangle%
\end{array}%
\right) =\frac{1}{\sqrt{2}}\left( 
\begin{array}{cc}
1 & 1 \\ 
1 & -1%
\end{array}%
\right) \left( 
\begin{array}{c}
\left\vert 0\right\rangle \\ 
\left\vert 1\right\rangle%
\end{array}%
\right) \text{.}
\end{equation}%
The unitary encoding operator $U_{\text{enc}}^{\text{phase}}$ is given by $%
U_{\text{enc}}^{\text{phase}}\overset{\text{def}}{=}H^{\otimes 3}\circ U_{%
\text{enc}}^{\text{bit}}$,with $U_{\text{enc}}^{\text{bit}}$ defined in (\ref%
{bitencoding}). Furthermore, in the phase flip code, the recovery operation
is the Hadamard conjugated recovery operation from the bit flip code, $%
R_{k}^{\text{phase}}\overset{\text{def}}{=}H^{\otimes 3}R_{k}^{\text{bit}%
}H^{\otimes 3}$.

\subsection{CASE, $n_{\text{even}}=4$}

Here, we apply the even length repetition code with $n=4$ to the correlated
bit-flip.

\emph{Error Operators}. In this case, the memory channel to consider is
given by,%
\begin{equation}
\Lambda ^{(4)}(\rho )\overset{\text{def}}{=}\sum_{i_{1}\text{, }i_{2}\text{, 
}i_{3}\text{, }%
i_{4}=0}^{1}p_{i_{4}|i_{3}}p_{i_{3}|i_{2}}p_{i_{2}|i_{1}}p_{i_{1}}\left[
A_{i_{4}}A_{i_{3}}A_{i_{2}}A_{i_{1}}\rho A_{i_{1}}^{\dag }A_{i_{2}}^{\dag
}A_{i_{3}}^{\dagger }A_{i_{4}}^{\dagger }\right] \text{, }  \label{n4}
\end{equation}%
where the error operators $\left\{ A_{i_{r}}\right\} $ with $r=1$, $2$, $3$, 
$4$ act on $1$ qubit quantum states. The error superoperator $\mathcal{A}$
associated to channel (\ref{n4}) may be defined in terms of the following
encoded error operators $\left\{ A_{k}^{\prime }\right\} $ with $k=0$,..,$15$%
,%
\begin{equation}
\mathcal{A}\longleftrightarrow \left\{ A_{0}^{\prime }\text{,.., }%
A_{15}^{\prime }\right\} \text{ with }\Lambda ^{(4)}(\rho )\overset{\text{def%
}}{=}\dsum\limits_{k=0}^{15}A_{k}^{\prime }\rho A_{k}^{\prime \dagger }\text{
and, }\dsum\limits_{k=0}^{15}A_{k}^{\prime \dagger }A_{k}^{\prime
}=I_{16\times 16}\text{.}
\end{equation}%
Omitting the symbol "$\otimes $", the error operators $\left\{ A_{0}^{\prime
}\text{,.., }A_{15}^{\prime }\right\} $ are given by,%
\begin{eqnarray}
A_{0}^{\prime } &=&\sqrt{\tilde{p}_{0}^{\left( 4\right) }}%
I^{1}I^{2}I^{3}I^{4}\text{, }A_{1}^{\prime }=\sqrt{\tilde{p}_{1}^{\left(
4\right) }}X^{1}I^{2}I^{3}I^{4}\text{, }A_{2}^{\prime }=\sqrt{\tilde{p}%
_{2}^{\left( 4\right) }}I^{1}X^{2}I^{3}I^{4}\text{, }A_{3}^{\prime }=\sqrt{%
\tilde{p}_{3}^{\left( 4\right) }}I^{1}I^{2}X^{3}I^{4}\text{, }A_{4}^{\prime
}=\sqrt{\tilde{p}_{4}^{\left( 4\right) }}I^{1}I^{2}I^{3}X^{4}\text{, } 
\notag \\
&&  \notag \\
A_{5}^{\prime } &=&\sqrt{\tilde{p}_{5}^{\left( 4\right) }}%
X^{1}X^{2}I^{3}I^{4}\text{, }A_{6}^{\prime }=\sqrt{\tilde{p}_{6}^{\left(
4\right) }}X^{1}I^{2}X^{3}I^{4}\text{, }A_{7}^{\prime }=\sqrt{\tilde{p}%
_{7}^{\left( 4\right) }}X^{1}I^{2}I^{3}X^{4}\text{, }A_{8}^{\prime }=\sqrt{%
\tilde{p}_{8}^{\left( 4\right) }}I^{1}X^{2}X^{3}I^{4}\text{, }A_{9}^{\prime
}=\sqrt{\tilde{p}_{9}^{\left( 4\right) }}I^{1}X^{2}I^{3}X^{4}\text{,}  \notag
\\
&&  \notag \\
A_{10}^{\prime } &=&\sqrt{\tilde{p}_{10}^{\left( 4\right) }}%
I^{1}I^{2}X^{3}X^{4}\text{, }A_{11}^{\prime }=\sqrt{\tilde{p}_{11}^{\left(
4\right) }}X^{1}X^{2}X^{3}I^{4}\text{, }A_{12}^{\prime }=\sqrt{\tilde{p}%
_{12}^{\left( 4\right) }}X^{1}X^{2}I^{3}X^{4}\text{, }A_{13}^{\prime }=\sqrt{%
\tilde{p}_{13}^{\left( 4\right) }}X^{1}I^{2}X^{3}X^{4}  \notag \\
&&  \notag \\
A_{14}^{\prime } &=&\sqrt{\tilde{p}_{14}^{\left( 4\right) }}%
I^{1}X^{2}X^{3}X^{4}\text{, }A_{15}^{\prime }=\sqrt{\tilde{p}_{15}^{\left(
4\right) }}X^{1}X^{2}X^{3}X^{4}\text{.}
\end{eqnarray}%
The coefficients $\tilde{p}_{k}^{\left( 4\right) }$ for $k=1$,.., $15$ are
formally given by $\tilde{p}_{k}^{\left( 4\right) }=p_{i_{1}^{\left(
4\right) }i_{2}^{\left( 4\right) }}p_{i_{2}^{\left( 4\right) }i_{3}^{\left(
4\right) }}p_{i_{3}^{\left( 4\right) }i_{4}^{\left( 4\right)
}}p_{i_{4}^{\left( 4\right) }}$ where the $i_{j}^{\left( 4\right) }\in
\left\{ 0\text{, }1\right\} $ are determined by the relation $A_{k}^{\prime
}\left\vert 0_{L}\right\rangle =\left\vert i_{1}^{\left( 4\right)
}i_{2}^{\left( 4\right) }i_{3}^{\left( 4\right) }i_{4}^{\left( 4\right)
}\right\rangle $. Following the line of reasoning used for the odd
repetition codes and omitting technical details that will appear in Appendix
A, the entanglement fidelity $\mathcal{F}_{RC}^{\left( 4\right) }\left( \mu 
\text{, }p\right) $ becomes,%
\begin{equation}
\mathcal{F}_{RC}^{\left( 4\right) }\left( \mu \text{, }p\right) =\mu
^{2}\left( 2p^{3}-3p^{2}+p\right) +\mu \left( -4p^{3}+6p^{2}-2p\right)
+\left( 2p^{3}-3p^{2}+1\right) \text{.}
\end{equation}%
Notice that $\mathcal{F}_{RC}^{\left( 4\right) }\left( \mu \text{, }p\right)
=\mathcal{F}_{RC}^{\left( 3\right) }\left( \mu \text{, }p\right) $ and, in
absence of correlations,%
\begin{equation}
\mathcal{F}_{RC}^{\left( 4\right) }\left( 0\text{, }p\right) =\sum_{m=0}^{1}%
\binom{4}{m}p^{m}\left( 1-p\right) ^{4-m}+\frac{1}{2}\left( 
\begin{array}{c}
4 \\ 
2%
\end{array}%
\right) p^{2}\left( 1-p\right) ^{2}=2p^{3}-3p^{2}+1\equiv \mathcal{F}%
_{RC}^{\left( 3\right) }\left( 0\text{, }p\right) \text{.}
\end{equation}%
Finally, it can also be shown that $\mathcal{F}_{RC}^{\left( 6\right)
}\left( \mu \text{, }p\right) =\mathcal{F}_{RC}^{\left( 5\right) }\left( \mu 
\text{, }p\right) $ with,%
\begin{eqnarray}
\mathcal{F}_{RC}^{\left( 5\right) }\left( \mu \text{, }p\right)  &=&\mu
^{4}\left( -6p^{5}+15p^{4}-12p^{3}+3p^{2}\right) +\mu ^{3}\left(
24p^{5}-60p^{4}+52p^{3}-18p^{2}+2p\right) +  \notag \\
&&  \notag \\
&&+\mu ^{2}\left( -36p^{5}+90p^{4}-78p^{3}+27p^{2}-3p\right) +\mu \left(
24p^{5}-60p^{4}+48p^{3}-12p^{2}\right) +  \notag \\
&&  \notag \\
&&+\left( -6p^{5}+15p^{4}-10p^{3}+1\right) \text{,}
\end{eqnarray}%
and $\mathcal{F}_{RC}^{\left( 8\right) }\left( \mu \text{, }p\right) =%
\mathcal{F}_{RC}^{\left( 7\right) }\left( \mu \text{, }p\right) $ with,%
\begin{eqnarray}
\mathcal{F}_{RC}^{\left( 7\right) }\left( \mu \text{, }p\right)  &=&\mu
^{6}\left( 20p^{7}-70p^{6}+90p^{5}-50p^{4}+10p^{3}\right) +\mu ^{5}\left(
-120p^{7}+420p^{6}-564p^{5}+360p^{4}-108p^{3}+12p^{2}\right) +  \notag \\
&&  \notag \\
&&+\mu ^{4}\left(
300p^{7}-1050p^{6}+1440p^{5}-975p^{4}+336p^{3}-54p^{2}+3p\right) +  \notag \\
&&  \notag \\
&&+\mu ^{3}\left(
-400p^{7}+1400p^{6}-1920p^{5}+1300p^{4}-448p^{3}+72p^{2}-4p\right) +  \notag
\\
&&  \notag \\
&&+\mu ^{2}\left(
300p^{7}-1050p^{6}+1410p^{5}-900p^{4}+270p^{3}-30p^{2}\right) +\mu \left(
-120p^{7}+420p^{6}-540p^{5}+300p^{4}-60p^{3}\right) +  \notag \\
&&  \notag \\
&&+\left( 20p^{7}-70p^{6}+84p^{5}-35p^{4}+1\right) \text{.}
\end{eqnarray}%
In Figure 1, we plot $\mathcal{F}_{RC}^{\left( 3\right) }\left( \mu \text{, }%
p\right) $, $\mathcal{F}_{RC}^{\left( 5\right) }\left( \mu \text{, }p\right) 
$ and $\mathcal{F}_{RC}^{\left( 7\right) }\left( \mu \text{, }p\right) $ vs. 
$\mu $ for $p=0.45$. From this plot, it is clear that the entanglement
fidelity $\mathcal{F}_{RC}^{\left( n\right) }\left( \mu \text{, }p\right) $
increases with increasing\ $n$ and decreases with the correlation parameter $%
\mu $. 

\section{Decoherence Free Subspaces for Correlated Bit Flip}

In this Section, we tackle our decoherence problem via the decoherence-free
subspaces formalism. This is a passive quantum error correction method where
the key idea is that of avoiding decoherence by encoding quantum information
into special subspaces that are protected from the interaction with the
environment by virtue of some specific dynamical symmetry. For a detailed
review, we refer to \cite{lidar03}.

\subsection{CASE, $n_{\text{odd}}=3$}

Let us consider the correlated bit-flip noisy error model as defined in (\ref%
{bit}) and (\ref{nota}).

\emph{Encoding and Decoding Operators}. Consider the following quantum code
encoding $1$ logical qubit into $3$-physical qubits,%
\begin{eqnarray}
\left\vert 0\right\rangle &\rightarrow &\left\vert 0_{L}\right\rangle 
\overset{\text{def}}{=}\frac{1}{\left( \sqrt{2}\right) ^{3}}\left(
\left\vert 0\right\rangle _{1}+\left\vert 1\right\rangle _{1}\right) \otimes
\left( \left\vert 0\right\rangle _{2}+\left\vert 1\right\rangle _{2}\right)
\otimes \left( \left\vert 0\right\rangle _{3}+\left\vert 1\right\rangle
_{3}\right) \equiv \left\vert +++\right\rangle \text{,}  \notag \\
&&  \notag \\
\left\vert 1\right\rangle &\rightarrow &\left\vert 1_{L}\right\rangle 
\overset{\text{def}}{=}\frac{1}{\left( \sqrt{2}\right) ^{3}}\left(
\left\vert 0\right\rangle _{1}-\left\vert 1\right\rangle _{1}\right) \otimes
\left( \left\vert 0\right\rangle _{2}-\left\vert 1\right\rangle _{2}\right)
\otimes \left( \left\vert 0\right\rangle _{3}-\left\vert 1\right\rangle
_{3}\right) \equiv \left\vert ---\right\rangle \text{,}  \label{placs2}
\end{eqnarray}%
with $\left\langle +++|+++\right\rangle =\left\langle ---|---\right\rangle
=1 $ and $\left\langle ---|+++\right\rangle =\left\langle
+++|---\right\rangle =0$.

\emph{Recovery Operators}. The set of error operators satisfying the
detectability condition $P_{\mathcal{C}}A_{k}^{\prime }P_{\mathcal{C}%
}=\lambda _{A_{k}^{\prime }}P_{\mathcal{C}}$ where $P_{\mathcal{C}%
}=\left\vert 0_{L}\right\rangle \left\langle 0_{L}\right\vert +$ $\left\vert
1_{L}\right\rangle \left\langle 1_{L}\right\vert $ is the projector operator
on the code subspace $\mathcal{C}=Span\left\{ \left\vert 0_{L}\right\rangle 
\text{, }\left\vert 1_{L}\right\rangle \right\} $ is given by,%
\begin{equation}
\mathcal{A}_{\text{detectable}}=\left\{ A_{0}^{\prime }\text{, }%
A_{4}^{\prime }\text{, }A_{5}^{\prime }\text{, }A_{6}^{\prime }\right\}
\subseteq \mathcal{A}\text{.}
\end{equation}%
Furthermore, since all the detectable errors are invertible, the set of
correctable errors is such that $\mathcal{A}_{\text{correctable}}^{\dagger }%
\mathcal{A}_{\text{correctable}}$ is detectable. It follows then that,%
\begin{equation}
\mathcal{A}_{\text{correctable}}=\left\{ A_{0}^{\prime }\text{, }%
A_{4}^{\prime }\text{, }A_{5}^{\prime }\text{, }A_{6}^{\prime }\right\}
\equiv \mathcal{A}_{\text{detectable}}\text{.}
\end{equation}%
The action of the correctable error operators $\mathcal{A}_{\text{correctable%
}}$ on the codewords $\left\vert 0_{L}\right\rangle $ and $\left\vert
1_{L}\right\rangle $ is given by,%
\begin{equation}
\left\vert 0_{L}\right\rangle \rightarrow A_{r}^{\prime }\left\vert
0_{L}\right\rangle =\sqrt{\tilde{p}_{r}^{\left( 3\right) }}\left\vert
0_{L}\right\rangle \text{, }A_{r}^{\prime }\left\vert 1_{L}\right\rangle =%
\sqrt{\tilde{p}_{r}^{\left( 3\right) }}\left\vert 1_{L}\right\rangle \text{,}
\label{dfsc}
\end{equation}%
for $r=0$, $4$, $5$, $6$. From (\ref{dfsc}), it follows that $\mathcal{C}%
=Span\left\{ \left\vert 0_{L}\right\rangle \text{, }\left\vert
1_{L}\right\rangle \right\} $ is a decoherence-free subspace for the
correctable error operators in $\mathcal{A}_{\text{correctable}}$. The two \
one-dimensional orthogonal subspaces $\mathcal{V}^{0_{L}}$ and $\mathcal{V}%
^{1_{L}}$ of $\mathcal{H}_{2}^{3}$ generated by the action of $\mathcal{A}_{%
\text{correctable}}$ on $\left\vert 0_{L}\right\rangle $ and $\left\vert
1_{L}\right\rangle $ are given by,%
\begin{equation}
\mathcal{V}^{0_{L}}=Span\left\{ \left\vert v_{1}^{0_{L}}\right\rangle
=\left\vert +++\right\rangle \right\} \text{, }
\end{equation}%
and,%
\begin{equation}
\mathcal{V}^{1_{L}}=Span\left\{ \left\vert v_{1}^{1_{L}}\right\rangle
=\left\vert ---\right\rangle \right\} \text{,}
\end{equation}%
respectively. Notice that $\mathcal{V}^{0_{L}}\oplus \mathcal{V}^{1_{L}}\neq 
\mathcal{H}_{2}^{3}$. This means that the trace preserving recovery
superoperator $\mathcal{R}$ is defined in terms of one standard recovery
operator $R_{1}$ and by the projector $R_{\perp }$ onto the orthogonal
complement of $\dbigoplus\limits_{i=0}^{1}\ \mathcal{V}^{i_{L}}$, i. e. the
part of the Hilbert space $\mathcal{H}_{2}^{3}$ which is not reached by
acting on the code $\mathcal{C}\ $\ with the correctable error operators. In
the case under consideration,%
\begin{equation}
R_{1}\overset{\text{def}}{=}\left\vert +++\right\rangle \left\langle
+++\right\vert +\left\vert ---\right\rangle \left\langle ---\right\vert 
\text{, }R_{\perp }=\sum_{s=1}^{6}\left\vert r_{s}\right\rangle \left\langle
r_{s}\right\vert \text{,}  \label{dfsr}
\end{equation}%
where $\left\{ \left\vert r_{s}\right\rangle \right\} $ is an orthonormal
basis for $\left( \mathcal{V}^{0_{L}}\oplus \mathcal{V}^{1_{L}}\right)
^{\perp }$. A suitable basis $\mathcal{B}_{\left( \mathcal{V}^{0_{L}}\oplus 
\mathcal{V}^{1_{L}}\right) ^{\perp }}$ is given by,%
\begin{equation}
\mathcal{B}_{\left( \mathcal{V}^{0_{L}}\oplus \mathcal{V}^{1_{L}}\right)
^{\perp }}=\left\{ r_{1}=\left\vert -++\right\rangle \text{, }%
r_{2}=\left\vert +-+\right\rangle \text{, }r_{3}=\left\vert ++-\right\rangle 
\text{, }r_{4}=\left\vert --+\right\rangle \text{, }r_{5}=\left\vert
-+-\right\rangle \text{, }r_{6}=\left\vert +--\right\rangle \right\} \text{.}
\end{equation}%
Therefore, $\mathcal{R}\leftrightarrow \left\{ R_{1}\text{, }R_{\perp
}\right\} $ is indeed a trace preserving quantum operation,%
\begin{equation}
R_{1}^{\dagger }R_{1}+R_{\perp }^{\dagger }R_{\perp }=I_{8\times 8}\text{.}
\end{equation}%
Considering this recovery operation $\mathcal{R}$ with $R_{2}\equiv R_{\perp
}$, the map $\Lambda ^{\left( 3\right) }\left( \rho \right) $ in (\ref{nota}%
) becomes,%
\begin{equation}
\Lambda _{\text{recover}}^{\left( 3\right) }\left( \rho \right) \equiv
\left( \mathcal{R\circ }\Lambda ^{(3)}\right) \left( \rho \right) \overset{%
\text{def}}{=}\sum_{k=0}^{7}\dsum\limits_{l=1}^{2}\left( R_{l}A_{k}^{\prime
}\right) \rho \left( R_{l}A_{k}^{\prime }\right) ^{\dagger }\text{,}
\label{pla4}
\end{equation}

\emph{Entanglement Fidelity}. We want to describe the action of $\mathcal{%
R\circ }\Lambda ^{(3)}$ restricted to the code subspace $\mathcal{C}$.
Therefore, we compute the $2\times 2$ matrix representation $\left[
R_{l}A_{k}^{\prime }\right] _{|\mathcal{C}}$ of each $R_{l}A_{k}^{\prime }$
with $l=1$, $2$ and $k=0$,.., $7$ where,%
\begin{equation}
\left[ R_{l}A_{k}^{\prime }\right] _{|\mathcal{C}}\overset{\text{def}}{=}%
\left( 
\begin{array}{cc}
\left\langle 0_{L}|R_{l}A_{k}^{\prime }|0_{L}\right\rangle & \left\langle
0_{L}|R_{l}A_{k}^{\prime }|1_{L}\right\rangle \\ 
\left\langle 1_{L}|R_{l}A_{k}^{\prime }|0_{L}\right\rangle & \left\langle
1_{L}|R_{l}A_{k}^{\prime }|1_{L}\right\rangle%
\end{array}%
\right) \text{.}  \label{sup}
\end{equation}%
Substituting (\ref{dfsc}) and (\ref{dfsr}) into (\ref{sup}), it turns out
that the only matrices $\left[ R_{l}A_{k}^{\prime }\right] _{|\mathcal{C}}$
with non-vanishing trace are given by,%
\begin{equation}
\left[ R_{1}A_{r}^{\prime }\right] _{|\mathcal{C}}=\sqrt{\tilde{p}%
_{r}^{\left( 3\right) }}\left( 
\begin{array}{cc}
1 & 0 \\ 
0 & 1%
\end{array}%
\right) \text{,}
\end{equation}%
for $r=0$, $4$, $5$, $6$. Therefore, the entanglement fidelity $\mathcal{F}%
_{DFS}^{\left( 3\right) }\left( \mu \text{, }p\right) $ defined as,%
\begin{equation}
\mathcal{F}_{DFS}^{\left( 3\right) }\left( \mu \text{, }p\right) \overset{%
\text{def}}{=}\mathcal{F}^{\left( 3\right) }\left( \frac{1}{2}I_{2\times 2}%
\text{, }\mathcal{R\circ }\Lambda ^{(3)}\right) =\frac{1}{\left( 2\right)
^{2}}\sum_{k=0}^{7}\dsum\limits_{l=1}^{2}\left\vert \text{tr}\left( \left[
R_{l}A_{k}^{\prime }\right] _{|\mathcal{C}}\right) \right\vert ^{2}\text{,}
\label{pla3}
\end{equation}%
results,%
\begin{equation}
\mathcal{F}_{DFS}^{\left( 3\right) }\left( \mu \text{, }p\right) =\tilde{p}%
_{0}^{\left( 3\right) }+\tilde{p}_{4}^{\left( 3\right) }+\tilde{p}%
_{5}^{\left( 3\right) }+\tilde{p}_{6}^{\left( 3\right) }\text{.}
\label{dfsusa}
\end{equation}%
The expression for $\mathcal{F}_{DFS}^{\left( 3\right) }\left( \mu \text{, }%
p\right) $ in (\ref{pla3}) represents the entanglement fidelity quantifying
the performance of the error correction scheme provided by the noiseless
code here considered. The quantum operation $\mathcal{R\circ }\Lambda ^{(3)}$
appearing in (\ref{pla3}) is defined in equation (\ref{pla4}) and the
recovery operators $R_{l}$ are explicitly given in (\ref{dfsr}). The action
of\textbf{\ }$R_{l}A_{k}^{\prime }$ in (\ref{pla3}) is restricted to the
code space $\mathcal{C}$ defined in (\ref{placs2}).

Substituting (\ref{usa2}) into (\ref{dfsusa}), we finally obtain%
\begin{equation}
\mathcal{F}_{DFS}^{\left( 3\right) }\left( \mu \text{, }p\right) =\mu
^{2}\left( -4p^{3}+6p^{2}-2p\right) +\mu \left( 8p^{3}-12p^{2}+4p\right)
+\left( -4p^{3}+6p^{2}-3p+1\right) \text{.}
\end{equation}

\subsection{CASE, $n_{\text{even}}=4$}

Let us now consider the correlated bit-flip noisy error model as defined in (%
\ref{n4}).

\emph{Encoding and Decoding Operators}. Consider the following quantum code
encoding $1$ logical qubit into $4$-physical qubits,%
\begin{eqnarray}
\left\vert 0\right\rangle  &\rightarrow &\left\vert 0_{L}\right\rangle 
\overset{\text{def}}{=}\frac{1}{\left( \sqrt{2}\right) ^{4}}\left(
\left\vert 0\right\rangle _{1}+\left\vert 1\right\rangle _{1}\right) \otimes
\left( \left\vert 0\right\rangle _{2}+\left\vert 1\right\rangle _{2}\right)
\otimes \left( \left\vert 0\right\rangle _{3}+\left\vert 1\right\rangle
_{3}\right) \otimes \left( \left\vert 0\right\rangle _{4}+\left\vert
1\right\rangle _{4}\right) \equiv \left\vert ++++\right\rangle \text{,} 
\notag \\
&&  \notag \\
\left\vert 1\right\rangle  &\rightarrow &\left\vert 1_{L}\right\rangle 
\overset{\text{def}}{=}\frac{1}{\left( \sqrt{2}\right) ^{4}}\left(
\left\vert 0\right\rangle _{1}-\left\vert 1\right\rangle _{1}\right) \otimes
\left( \left\vert 0\right\rangle _{2}-\left\vert 1\right\rangle _{2}\right)
\otimes \left( \left\vert 0\right\rangle _{3}-\left\vert 1\right\rangle
_{3}\right) \otimes \left( \left\vert 0\right\rangle _{4}+\left\vert
1\right\rangle _{4}\right) \equiv \left\vert ----\right\rangle \text{,}
\end{eqnarray}%
with $\left\langle ++++|++++\right\rangle =\left\langle
----|----\right\rangle =1$ and $\left\langle ----|++++\right\rangle
=\left\langle ++++|----\right\rangle =0$. Following the line of reasoning
used for the odd case and omitting technical details that will appear in
Appendix B, the entanglement fidelity $\mathcal{F}_{DFS}^{\left( 4\right)
}\left( \mu \text{, }p\right) $ becomes,%
\begin{eqnarray}
\mathcal{F}_{DFS}^{\left( 4\right) }\left( \mu \text{, }p\right)  &=&\mu
^{3}\left( -8p^{4}+16p^{3}-10p^{2}+2p\right) +\mu ^{2}\left(
24p^{4}-48p^{3}+28p^{2}-4p\right) +  \notag \\
&&  \notag \\
&&+\mu \left( -24p^{4}+48p^{3}-30p^{2}+6p\right) +\left(
8p^{4}-16p^{3}+12p^{2}-4p+1\right) \text{.}
\end{eqnarray}%
In Figure 2, we plot $\mathcal{F}_{RC}^{\left( 4\right) }\left( \mu \text{, }%
p\right) $ and $\mathcal{F}_{DFS}^{\left( 4\right) }\left( \mu \text{, }%
p\right) $ for three values of the error probability $p=0.45$, $p=0.40$ and $%
p=0.35$. Following the line of reasoning presented above, it can be shown
that $\mathcal{F}_{DFS}^{\left( 5\right) }\left( \mu \text{, }p\right) $ and 
$\mathcal{F}_{DFS}^{\left( 6\right) }\left( \mu \text{, }p\right) $ are
given by,%
\begin{eqnarray}
\mathcal{F}_{DFS}^{\left( 5\right) }\left( \mu \text{, }p\right)  &=&\mu
^{4}\left( -16p^{5}+40p^{4}-36p^{3}+14p^{2}-2p\right) +\mu ^{3}\left(
64p^{5}-160p^{4}+136p^{3}-44p^{2}+4p\right) +  \notag \\
&&  \notag \\
&&+\mu ^{2}\left( -96p^{5}+240p^{4}-204p^{3}+66p^{2}-6p\right) +\mu \left(
64p^{5}-160p^{4}\allowbreak +144\allowbreak p^{3}-56\allowbreak
p^{2}+8p\right) +  \notag \\
&&  \notag \\
&&+\left( -16\allowbreak p^{5}+40p^{4}-40p^{3}+20p^{2}-5p+\allowbreak
1\right) \text{,}
\end{eqnarray}%
and,%
\begin{eqnarray}
\mathcal{F}_{DFS}^{\left( 6\right) }\left( \mu \text{, }p\right)  &=&\mu
^{5}\left( -32p^{6}+97p^{5}-115p^{4}+67p^{3}-19p^{2}+2p\right) +  \notag \\
&&  \notag \\
&&+\mu ^{4}\left( 160p^{6}-484p^{5}+546p^{4}-280p^{3}+62p^{2}-4p\right) + 
\notag \\
&&  \notag \\
&&+\mu ^{3}\left( -320p^{6}+966p^{5}-1068p^{4}+519p^{3}-103p^{2}+6p\right) +
\notag \\
&&  \notag \\
&&+\mu ^{2}\left( 320p^{6}-964p^{5}+1078p^{4}-546p^{3}+120p^{2}-8p\right) + 
\notag \\
&&  \notag \\
&&+\mu \left( -160p^{6}+481p^{5}-561p^{4}+320p^{3}-90p^{2}+10p\right) + 
\notag \\
&&  \notag \\
&&+\left( 32p^{6}-96p^{5}+120p^{4}-80p^{3}+30p^{2}-6p+1\right) \text{.}
\end{eqnarray}%
respectively. It turns out that $\mathcal{F}_{DFS}^{\left( 5\right) }\left(
\mu \text{, }p\right) \leq \mathcal{F}_{DFS}^{\left( 3\right) }\left( \mu 
\text{, }p\right) $ and $\mathcal{F}_{DFS}^{\left( 6\right) }\left( \mu 
\text{, }p\right) \leq \mathcal{F}_{DFS}^{\left( 4\right) }\left( \mu \text{%
, }p\right) $ for $\mu \in \left[ 0\text{, }1\right] $ and $p<0.5$.
Moreover, $\mathcal{F}_{DFS}^{\left( 4\right) }\left( \mu \text{, }p\right) $
is greater than $\mathcal{F}_{DFS}^{\left( 3\right) }\left( \mu \text{, }%
p\right) $ for $\mu \geq \mu _{\text{min}}\geq 0.2$ . Therefore, in the high
correlation regime where $\mu \rightarrow 1$, $\mathcal{F}_{\text{DFS}%
}^{\left( 4\right) }\left( \mu \text{, }p\right) $ achieves the highest
value for arbitrary error probabilities $p$ less than $0.5$.

\section{Final Remarks}

Because of the results obtained in the previous Section, it follows that
there must be a certain threshold value $\mu ^{\ast }\left( p\right) $ that
allows to select the better code between the repetition and the noiseless
quantum code for our noisy quantum memory channel. Considering the case with 
$n=4$, we may obtain a curve $\mu ^{\ast }=\mu ^{\ast }\left( p\right) $
defined in such a way that, $\mathcal{F}_{DFS}^{\left( 4\right) }\left( \mu
^{\ast }\left( p\right) \text{, }p\right) -\mathcal{F}_{RC}^{\left( 4\right)
}\left( \mu ^{\ast }\left( p\right) \text{, }p\right) =0$. For example, In
Figure 2 we have plotted $\mathcal{F}_{RC}^{\left( 4\right) }\left( \mu 
\text{, }p\right) $ and $\mathcal{F}_{DFS}^{\left( 4\right) }\left( \mu 
\text{, }p\right) $ for few values of $p$. From this plot, we see\textbf{\ }%
the emergence of threshold values $\mu ^{\ast }\left( 0.45\right) \simeq 0.34
$, $\mu ^{\ast }\left( 0.40\right) \simeq 0.45$ and $\mu ^{\ast }\left(
0.35\right) \simeq 0.52$ when the curves $\mathcal{F}_{RC}^{\left( 4\right)
}\left( \mu \text{, }p\right) $\ and $\mathcal{F}_{DFS}^{\left( 4\right)
}\left( \mu \text{, }p\right) $ cross. This means that for $p=0.45$ the
noiseless quantum code outperforms the repetition code when $\mu \geq \mu
^{\ast }\left( 0.45\right) \simeq 0.34$. Finally, in Figure 3 we plot the
threshold curve $\mu ^{\ast }\left( p\right) $ for all permitted values of
the error probability $p$.\ In conclusion, we have shown in an explicit way
that the repetition code (be it even or odd) works better than the noiseless
quantum code in the low correlations regime. On the contrary, in the high
correlation regime, the noiseless quantum codes work better. The proper
quantities defining the correlation regimes are the threshold values $\mu
^{\ast }\left( p\right) $.

In conclusion, in this Letter we have analyzed the performance of simple
quantum error correcting codes in the presence of correlated noise error
models characterized by a correlation strength $\mu $. Specifically, we have
considered bit flip (phase flip) noisy quantum memory channels and used
repetition and noiseless quantum codes. We have characterized the
performance of the codes by means of the entanglement fidelity $\mathcal{F}%
\left( \mu \text{, }p\right) $ as function of the error probability $p$ and
degree of memory $\mu $. We have shown in an explicit way that the
entanglement fidelity $\mathcal{F}_{RC}^{\left( n_{odd}\right) }\left( \mu 
\text{, }p\right) $ equals $\mathcal{F}_{RC}^{\left( n_{odd}+1\right)
}\left( \mu \text{, }p\right) $ and that $\mathcal{F}_{RC}^{\left( n\right)
}\left( \mu \text{, }p\right) $ increases with the length $n$ of the code
and decreases with the correlation parameter $\mu $. Furthermore, we also
used the decoherence free subspaces formalism and showed that the
performance of such QECCs quantified in terms of the entanglement fidelity $%
\mathcal{F}_{DFS}^{\left( n\right) }\left( \mu \text{, }p\right) $ is better
than the one of repetition codes in the high correlation regime where $\mu
\rightarrow 1$. The noiseless quantum code with $n=4$ preforms better than
the other (noiseless) codes considered in this work in the high correlation
regime. Comparing the entanglement fidelities of repetition codes and
noiseless quantum codes, we found a threshold $\mu ^{\ast }\left( p\right) $
for the correlation strength that allows to select the quantum code with
better performance.

The above results suggest that it may be convenient to concatenate
decoherence-free subspaces with standard quantum error correcting codes in
order to achieve higher entanglement fidelity values in both low and high
correlations regimes. This will be the object of future investigations.

\begin{acknowledgments}
C. C. thanks C. Lupo and L. Memarzadeh for very useful discussions. This
work was supported by the European Community's Seventh Framework Program
(CORNER Project; FP7/2007-2013) under grant agreement 213681.
\end{acknowledgments}

\appendix

\section{Repetition Codes, $n_{\text{even}}=4$}

\emph{Recovery Operators}. The set of error operators satisfying the
detectability condition $P_{\mathcal{C}}A_{k}^{\prime }P_{\mathcal{C}%
}=\lambda _{A_{k}^{\prime }}P_{\mathcal{C}}$ where $P_{\mathcal{C}%
}=\left\vert 0_{L}\right\rangle \left\langle 0_{L}\right\vert +$ $\left\vert
1_{L}\right\rangle \left\langle 1_{L}\right\vert $ is the projector operator
on the code subspace $\mathcal{C}=Span\left\{ \left\vert 0_{L}\right\rangle 
\text{, }\left\vert 1_{L}\right\rangle \right\} $ with $\left\vert
0_{L}\right\rangle \overset{\text{def}}{=}\left\vert 0000\right\rangle $ and 
$\left\vert 1_{L}\right\rangle \overset{\text{def}}{=}\left\vert
1111\right\rangle $ is given by $\mathcal{A}_{\text{detectable}}=\mathcal{A}%
\backslash \left\{ A_{15}^{\prime }\right\} \subseteq \mathcal{A}$.
Furthermore, since all the detectable errors are invertible, the set of
correctable errors is such that $\mathcal{A}_{\text{correctable}}^{\dagger }%
\mathcal{A}_{\text{correctable}}$ is detectable. It follows then that,%
\begin{equation}
\mathcal{A}_{\text{correctable}}=\left\{ A_{0}^{\prime }\text{, }%
A_{1}^{\prime }\text{, }A_{2}^{\prime }\text{, }A_{3}^{\prime }\text{, }%
A_{4}^{\prime }\text{, }A_{5}^{\prime }\text{, }A_{6}^{\prime }\text{, }%
A_{7}^{\prime }\right\} \subseteq \mathcal{A}_{\text{detectable}}\subseteq 
\mathcal{A}\text{.}
\end{equation}%
The action of the correctable error operators $\mathcal{A}_{\text{correctable%
}}$ on the codewords $\left\vert 0_{L}\right\rangle $ and $\left\vert
1_{L}\right\rangle $ is given by,%
\begin{eqnarray}
\left\vert 0_{L}\right\rangle &\rightarrow &A_{0}^{\prime }\left\vert
0_{L}\right\rangle =\sqrt{\tilde{p}_{0}^{\left( 4\right) }}\left\vert
0000\right\rangle \text{, }A_{1}^{\prime }\left\vert 0_{L}\right\rangle =%
\sqrt{\tilde{p}_{1}^{\left( 4\right) }}\left\vert 1000\right\rangle \text{, }%
A_{2}^{\prime }\left\vert 0_{L}\right\rangle =\sqrt{\tilde{p}_{2}^{\left(
4\right) }}\left\vert 0100\right\rangle \text{, }A_{3}^{\prime }\left\vert
0_{L}\right\rangle =\sqrt{\tilde{p}_{3}^{\left( 4\right) }}\left\vert
0010\right\rangle \text{, }  \notag \\
&&  \notag \\
A_{4}^{\prime }\left\vert 0_{L}\right\rangle &=&\sqrt{\tilde{p}_{4}^{\left(
4\right) }}\left\vert 0001\right\rangle \text{, }A_{5}^{\prime }\left\vert
0_{L}\right\rangle =\sqrt{\tilde{p}_{5}^{\left( 4\right) }}\left\vert
1100\right\rangle \text{, }A_{6}^{\prime }\left\vert 0_{L}\right\rangle =%
\sqrt{\tilde{p}_{6}^{\left( 4\right) }}\left\vert 1010\right\rangle \text{, }%
A_{7}^{\prime }\left\vert 0_{L}\right\rangle =\sqrt{\tilde{p}_{7}^{\left(
4\right) }}\left\vert 1001\right\rangle \text{,}  \label{ea1}
\end{eqnarray}%
and,%
\begin{eqnarray}
\left\vert 1_{L}\right\rangle &\rightarrow &A_{0}^{\prime }\left\vert
1_{L}\right\rangle =\sqrt{\tilde{p}_{0}^{\left( 4\right) }}\left\vert
1_{L}\right\rangle \text{, }A_{1}^{\prime }\left\vert 1_{L}\right\rangle =%
\sqrt{\tilde{p}_{1}^{\left( 4\right) }}\left\vert 0111\right\rangle \text{, }%
A_{2}^{\prime }\left\vert 1_{L}\right\rangle =\sqrt{\tilde{p}_{2}^{\left(
4\right) }}\left\vert 1011\right\rangle \text{, }A_{3}^{\prime }\left\vert
1_{L}\right\rangle =\sqrt{\tilde{p}_{3}^{\left( 4\right) }}\left\vert
1101\right\rangle \text{,}  \notag \\
&&  \notag \\
A_{4}^{\prime }\left\vert 1_{L}\right\rangle &=&\sqrt{\tilde{p}_{4}^{\left(
4\right) }}\left\vert 1110\right\rangle \text{, }A_{5}^{\prime }\left\vert
1_{L}\right\rangle =\sqrt{\tilde{p}_{5}^{\left( 4\right) }}\left\vert
0011\right\rangle \text{, }A_{6}^{\prime }\left\vert 1_{L}\right\rangle =%
\sqrt{\tilde{p}_{6}^{\left( 4\right) }}\left\vert 0101\right\rangle \text{, }%
A_{7}^{\prime }\left\vert 1_{L}\right\rangle =\sqrt{\tilde{p}_{7}^{\left(
4\right) }}\left\vert 0110\right\rangle \text{,}  \label{ea2}
\end{eqnarray}%
respectively. The two \ eight-dimensional orthogonal subspaces $\mathcal{V}%
^{0_{L}}$ and $\mathcal{V}^{1_{L}}$ of $\mathcal{H}_{2}^{4}$ generated by
the action of $\mathcal{A}_{\text{correctable}}$ on $\left\vert
0_{L}\right\rangle $ and $\left\vert 1_{L}\right\rangle $ are given by,%
\begin{equation}
\mathcal{V}^{0_{L}}=Span\left\{ 
\begin{array}{c}
\left\vert v_{1}^{0_{L}}\right\rangle =\left\vert 0000\right\rangle \text{,}%
\left\vert v_{2}^{0_{L}}\right\rangle =\left\vert 1000\right\rangle \text{, }%
\left\vert v_{3}^{0_{L}}\right\rangle =\left\vert 0100\right\rangle \text{, }%
\left\vert v_{4}^{0_{L}}\right\rangle =\left\vert 0010\right\rangle \text{,}
\\ 
\\ 
\text{ }\left\vert v_{5}^{0_{L}}\right\rangle =\left\vert 0001\right\rangle 
\text{,}\left\vert v_{6}^{0_{L}}\right\rangle =\left\vert 1100\right\rangle 
\text{, }\left\vert v_{7}^{0_{L}}\right\rangle =\left\vert 1010\right\rangle 
\text{, }\left\vert v_{8}^{0_{L}}\right\rangle =\left\vert 1001\right\rangle 
\text{,}%
\end{array}%
\right\} \text{,}  \label{span11}
\end{equation}%
and,%
\begin{equation}
\mathcal{V}^{1_{L}}=Span\left\{ 
\begin{array}{c}
\left\vert v_{1}^{1_{L}}\right\rangle =\left\vert 1111\right\rangle \text{,}%
\left\vert v_{2}^{1_{L}}\right\rangle =\left\vert 0111\right\rangle \text{, }%
\left\vert v_{3}^{1_{L}}\right\rangle =\left\vert 1011\right\rangle \text{, }%
\left\vert v_{4}^{1_{L}}\right\rangle =\left\vert 1101\right\rangle \text{,}
\\ 
\\ 
\left\vert v_{5}^{1_{L}}\right\rangle =\left\vert 1110\right\rangle \text{,}%
\left\vert v_{6}^{1_{L}}\right\rangle =\left\vert 0011\right\rangle \text{, }%
\left\vert v_{7}^{1_{L}}\right\rangle =\left\vert 0101\right\rangle \text{, }%
\left\vert v_{8}^{1_{L}}\right\rangle =\left\vert 0110\right\rangle%
\end{array}%
\right\} \text{.}  \label{span21}
\end{equation}%
Notice that $\mathcal{V}^{0_{L}}\oplus \mathcal{V}^{1_{L}}=\mathcal{H}%
_{2}^{4}$. The recovery superoperator $\mathcal{R}\leftrightarrow \left\{
R_{l}\right\} $ with $l=1$,.., $8$ is defined as,%
\begin{equation}
R_{l}\overset{\text{def}}{=}V_{l}\sum_{i=0}^{1}\left\vert
v_{l}^{i_{L}}\right\rangle \left\langle v_{l}^{i_{L}}\right\vert \text{,}
\label{recovery1}
\end{equation}%
where the unitary operator $V_{l}$ is such that $V_{l}\left\vert
v_{l}^{i_{L}}\right\rangle =\left\vert i_{L}\right\rangle $ for $i\in
\left\{ 0\text{, }1\right\} $. Substituting (\ref{span11}) and (\ref{span21}%
) into (\ref{recovery1}), it follows that the eight recovery operators $%
\left\{ R_{1}\text{, .., }R_{8}\right\} $ are given by,%
\begin{eqnarray}
R_{1} &=&\left\vert 0_{L}\right\rangle \left\langle 0_{L}\right\vert
+\left\vert 1_{L}\right\rangle \left\langle 1_{L}\right\vert \text{, }%
R_{2}=\left\vert 0_{L}\right\rangle \left\langle 1000\right\vert +\left\vert
1_{L}\right\rangle \left\langle 0111\right\vert \text{, }R_{3}=\left\vert
0_{L}\right\rangle \left\langle 0100\right\vert +\left\vert
1_{L}\right\rangle \left\langle 1011\right\vert \text{,}  \notag \\
&&  \notag \\
\text{ }R_{4} &=&\left\vert 0_{L}\right\rangle \left\langle 0010\right\vert
+\left\vert 1_{L}\right\rangle \left\langle 1101\right\vert \text{, }%
R_{5}=\left\vert 0_{L}\right\rangle \left\langle 0001\right\vert +\left\vert
1_{L}\right\rangle \left\langle 1110\right\vert \text{, }R_{6}=\left\vert
0_{L}\right\rangle \left\langle 1100\right\vert +\left\vert
1_{L}\right\rangle \left\langle 0011\right\vert  \notag \\
&&  \notag \\
R_{7} &=&\left\vert 0_{L}\right\rangle \left\langle 1010\right\vert
+\left\vert 1_{L}\right\rangle \left\langle 0101\right\vert \text{, }%
R_{8}=\left\vert 0_{L}\right\rangle \left\langle 1001\right\vert +\left\vert
1_{L}\right\rangle \left\langle 0110\right\vert \text{.}  \label{rec1}
\end{eqnarray}%
It can be shown that $\mathcal{R}\leftrightarrow \left\{ R_{l}\right\} $
with $l=1$,.., $8$ is indeed a trace preserving quantum operation since,%
\begin{equation}
\sum_{l=1}^{8}R_{l}^{\dagger }R_{l}=I_{16\times 16}\text{.}
\end{equation}%
Considering this recovery operation $\mathcal{R}$, the map $\Lambda ^{\left(
4\right) }\left( \rho \right) $ in (\ref{n4}) becomes,%
\begin{equation}
\Lambda _{\text{recover}}^{\left( 4\right) }\left( \rho \right) \equiv
\left( \mathcal{R\circ }\Lambda ^{(4)}\right) \left( \rho \right) \overset{%
\text{def}}{=}\sum_{k=0}^{15}\dsum\limits_{l=1}^{8}\left( R_{l}A_{k}^{\prime
}\right) \rho \left( R_{l}A_{k}^{\prime }\right) ^{\dagger }\text{.}
\end{equation}

\emph{Entanglement Fidelity}. We want to describe the action of $\mathcal{%
R\circ }\Lambda ^{(4)}$ restricted to the code subspace $\mathcal{C}$. We
simply compute the $2\times 2$ matrix representation $\left[
R_{l}A_{k}^{\prime }\right] _{|\mathcal{C}}$ of each $R_{l}A_{k}^{\prime }$
with $l=1$,.., $4$ and $k=0$,.., $7$ where,%
\begin{equation}
\left[ R_{l}A_{k}^{\prime }\right] _{|\mathcal{C}}\overset{\text{def}}{=}%
\left( 
\begin{array}{cc}
\left\langle 0_{L}|R_{l}A_{k}^{\prime }|0_{L}\right\rangle & \left\langle
0_{L}|R_{l}A_{k}^{\prime }|1_{L}\right\rangle \\ 
\left\langle 1_{L}|R_{l}A_{k}^{\prime }|0_{L}\right\rangle & \left\langle
1_{L}|R_{l}A_{k}^{\prime }|1_{L}\right\rangle%
\end{array}%
\right) \text{.}  \label{sopra11}
\end{equation}%
Substituting (\ref{ea1}), (\ref{ea2}) and (\ref{rec1}) into (\ref{sopra11}),
it turns out that the only matrices $\left[ R_{l}A_{k}^{\prime }\right] _{|%
\mathcal{C}}$ with non-vanishing trace are given by,%
\begin{eqnarray}
\left[ R_{1}A_{0}^{\prime }\right] _{|\mathcal{C}} &=&\sqrt{\tilde{p}%
_{0}^{\left( 4\right) }}\left( 
\begin{array}{cc}
1 & 0 \\ 
0 & 1%
\end{array}%
\right) \text{, }\left[ R_{2}A_{1}^{\prime }\right] _{|\mathcal{C}}=\sqrt{%
\tilde{p}_{1}^{\left( 4\right) }}\left( 
\begin{array}{cc}
1 & 0 \\ 
0 & 1%
\end{array}%
\right) \text{, }\left[ R_{3}A_{2}^{\prime }\right] _{|\mathcal{C}}=\sqrt{%
\tilde{p}_{2}^{\left( 4\right) }}\left( 
\begin{array}{cc}
1 & 0 \\ 
0 & 1%
\end{array}%
\right) \text{, }  \notag \\
&&  \notag \\
\left[ R_{4}A_{3}^{\prime }\right] _{|\mathcal{C}} &=&\sqrt{\tilde{p}%
_{3}^{\left( 4\right) }}\left( 
\begin{array}{cc}
1 & 0 \\ 
0 & 1%
\end{array}%
\right) \text{, }\left[ R_{5}A_{4}^{\prime }\right] _{|\mathcal{C}}=\sqrt{%
\tilde{p}_{4}^{\left( 4\right) }}\left( 
\begin{array}{cc}
1 & 0 \\ 
0 & 1%
\end{array}%
\right) \text{, }\left[ R_{6}A_{5}^{\prime }\right] _{|\mathcal{C}}=\sqrt{%
\tilde{p}_{5}^{\left( 4\right) }}\left( 
\begin{array}{cc}
1 & 0 \\ 
0 & 1%
\end{array}%
\right) \text{,}  \notag \\
&&  \notag \\
\left[ R_{7}A_{6}^{\prime }\right] _{|\mathcal{C}} &=&\sqrt{\tilde{p}%
_{6}^{\left( 4\right) }}\left( 
\begin{array}{cc}
1 & 0 \\ 
0 & 1%
\end{array}%
\right) \text{, }\left[ R_{8}A_{7}^{\prime }\right] _{|\mathcal{C}}=\sqrt{%
\tilde{p}_{7}^{\left( 4\right) }}\left( 
\begin{array}{cc}
1 & 0 \\ 
0 & 1%
\end{array}%
\right) \text{.}
\end{eqnarray}%
Therefore, the entanglement fidelity $\mathcal{F}_{RC}^{\left( 4\right)
}\left( \mu \text{, }p\right) $ defined as,%
\begin{equation}
\mathcal{F}_{RC}^{\left( 4\right) }\left( \mu \text{, }p\right) \overset{%
\text{def}}{=}\mathcal{F}_{RC}^{\left( 4\right) }\left( \frac{1}{2}%
I_{2\times 2}\text{, }\mathcal{R\circ }\Lambda ^{(4)}\right) =\frac{1}{%
\left( 2\right) ^{2}}\sum_{k=0}^{15}\dsum\limits_{l=1}^{8}\left\vert \text{tr%
}\left( \left[ R_{l}A_{k}^{\prime }\right] _{|\mathcal{C}}\right)
\right\vert ^{2}\text{,}
\end{equation}%
results,%
\begin{equation}
\mathcal{F}_{RC}^{\left( 4\right) }\left( \mu \text{, }p\right) =\tilde{p}%
_{0}^{\left( 4\right) }+\tilde{p}_{1}^{\left( 4\right) }+\tilde{p}%
_{2}^{\left( 4\right) }+\tilde{p}_{3}^{\left( 4\right) }+\tilde{p}%
_{4}^{\left( 4\right) }+\tilde{p}_{5}^{\left( 4\right) }+\tilde{p}%
_{6}^{\left( 4\right) }+\tilde{p}_{7}^{\left( 4\right) }\text{.}
\label{usa31}
\end{equation}%
Substituting (\ref{usa2}) into (\ref{usa31}), we finally obtain%
\begin{equation}
\mathcal{F}_{RC}^{\left( 4\right) }\left( \mu \text{, }p\right) =\mu
^{2}\left( 2p^{3}-3p^{2}+p\right) +\mu \left( -4p^{3}+6p^{2}-2p\right)
+\left( 2p^{3}-3p^{2}+1\right) \text{.}
\end{equation}%
Notice that $\mathcal{F}_{RC}^{\left( 4\right) }\left( \mu \text{, }p\right)
=\mathcal{F}_{RC}^{\left( 3\right) }\left( \mu \text{, }p\right) $ and, in
absence of correlations,%
\begin{equation}
\mathcal{F}_{RC}^{\left( 4\right) }\left( 0\text{, }p\right) =\sum_{m=0}^{1}%
\binom{4}{m}p^{m}\left( 1-p\right) ^{4-m}+\frac{1}{2}\left( 
\begin{array}{c}
4 \\ 
2%
\end{array}%
\right) p^{2}\left( 1-p\right) ^{2}=2p^{3}-3p^{2}+1\equiv \mathcal{F}%
_{RC}^{\left( 3\right) }\left( 0\text{, }p\right) \text{.}
\end{equation}%
Finally, following the same line of reasoning presented above, it can be
shown that $\mathcal{F}_{RC}^{\left( 6\right) }\left( \mu \text{, }p\right) =%
\mathcal{F}_{RC}^{\left( 5\right) }\left( \mu \text{, }p\right) $ and $%
\mathcal{F}_{RC}^{\left( 8\right) }\left( \mu \text{, }p\right) =\mathcal{F}%
_{RC}^{\left( 7\right) }\left( \mu \text{, }p\right) $.

\section{Decoherence Free Subspaces, $n_{\text{even}}=4$}

\emph{Recovery Operators}. The set of error operators satisfying the
detectability condition $P_{\mathcal{C}}A_{k}^{\prime }P_{\mathcal{C}%
}=\lambda _{A_{k}^{\prime }}P_{\mathcal{C}}$ where $P_{\mathcal{C}%
}=\left\vert 0_{L}\right\rangle \left\langle 0_{L}\right\vert +$ $\left\vert
1_{L}\right\rangle \left\langle 1_{L}\right\vert $ is the projector operator
on the code subspace $\mathcal{C}=Span\left\{ \left\vert 0_{L}\right\rangle 
\text{, }\left\vert 1_{L}\right\rangle \right\} $ is given by,%
\begin{equation}
\mathcal{A}_{\text{detectable}}=\left\{ A_{0}^{\prime }\text{, }%
A_{5}^{\prime }\text{, }A_{6}^{\prime }\text{, }A_{7}^{\prime }\text{, }%
A_{8}^{\prime }\text{, }A_{9}^{\prime }\text{, }A_{10}^{\prime }\text{, }%
A_{15}^{\prime }\right\} \subseteq \mathcal{A}\text{.}
\end{equation}%
Furthermore, since all the detectable errors are invertible, the set of
correctable errors is such that $\mathcal{A}_{\text{correctable}}^{\dagger }%
\mathcal{A}_{\text{correctable}}$ is detectable. It follows then that,%
\begin{equation}
\mathcal{A}_{\text{correctable}}=\left\{ A_{0}^{\prime }\text{, }%
A_{5}^{\prime }\text{, }A_{6}^{\prime }\text{, }A_{7}^{\prime }\text{, }%
A_{8}^{\prime }\text{, }A_{9}^{\prime }\text{, }A_{10}^{\prime }\text{, }%
A_{15}^{\prime }\right\} \equiv \mathcal{A}_{\text{detectable}}\text{.}
\end{equation}%
The action of the correctable error operators $\mathcal{A}_{\text{correctable%
}}$ on the codewords $\left\vert 0_{L}\right\rangle $ and $\left\vert
1_{L}\right\rangle $ is given by,%
\begin{equation}
\left\vert 0_{L}\right\rangle \rightarrow A_{0r}^{\prime }\left\vert
0_{L}\right\rangle =\sqrt{\tilde{p}_{r}^{\left( 4\right) }}\left\vert
0_{L}\right\rangle \text{, }\left\vert 1_{L}\right\rangle \rightarrow
A_{r}^{\prime }\left\vert 1_{L}\right\rangle =\sqrt{\tilde{p}_{r}^{\left(
4\right) }}\left\vert 1_{L}\right\rangle \text{,}  \label{1dd}
\end{equation}%
for $r=0$, $5$, $6$, $7$, $8$, $9$, $10$, $15$. From (\ref{1dd}), it follows
that $\mathcal{C}=Span\left\{ \left\vert 0_{L}\right\rangle \text{, }%
\left\vert 1_{L}\right\rangle \right\} $ is a decoherence-free subspace for
the correctable error operators in $\mathcal{A}_{\text{correctable}}$. The
two \ one-dimensional orthogonal subspaces $\mathcal{V}^{0_{L}}$ and $%
\mathcal{V}^{1_{L}}$ of $\mathcal{H}_{2}^{4}$ generated by the action of $%
\mathcal{A}_{\text{correctable}}$ on $\left\vert 0_{L}\right\rangle $ and $%
\left\vert 1_{L}\right\rangle $ are given by,%
\begin{equation}
\mathcal{V}^{0_{L}}=Span\left\{ \left\vert v_{1}^{0_{L}}\right\rangle
=\left\vert ++++\right\rangle \right\} \text{,}
\end{equation}%
and,%
\begin{equation}
\mathcal{V}^{1_{L}}=Span\left\{ \left\vert v_{1}^{1_{L}}\right\rangle
=\left\vert ----\right\rangle \right\} \text{,}
\end{equation}%
respectively. Notice that $\mathcal{V}^{0_{L}}\oplus \mathcal{V}^{1_{L}}\neq 
\mathcal{H}_{2}^{4}$. This means that the trace preserving recovery
superoperator $\mathcal{R}$ is defined in terms of one standard recovery
operator $R_{1}$ and by the projector $R_{\perp }$ onto the orthogonal
complement of $\dbigoplus\limits_{i=0}^{1}\ \mathcal{V}^{i_{L}}$, i. e. the
part of the Hilbert space $\mathcal{H}_{2}^{4}$ which is not reached by
acting on the code $\mathcal{C}\ $\ with the correctable error operators. In
the case under consideration,%
\begin{equation}
R_{1}\overset{\text{def}}{=}\left\vert ++++\right\rangle \left\langle
++++\right\vert +\left\vert ----\right\rangle \left\langle ----\right\vert 
\text{, }R_{\perp }=\sum_{s=1}^{14}\left\vert r_{s}\right\rangle
\left\langle r_{s}\right\vert \text{,}
\end{equation}%
where $\left\{ \left\vert r_{s}\right\rangle \right\} $ is an orthonormal
basis for $\left( \mathcal{V}^{0_{L}}\oplus \mathcal{V}^{1_{L}}\right)
^{\perp }$. It can be shown that $\mathcal{R}\leftrightarrow \left\{ R_{1}%
\text{, }R_{\perp }\right\} $ is a trace preserving quantum operation,%
\begin{equation}
R_{1}^{\dagger }R_{1}+R_{\perp }^{\dagger }R_{\perp }=I_{16\times 16}\text{.}
\end{equation}%
Considering this recovery operation $\mathcal{R}$ with $R_{2}\equiv R_{\perp
}$, the map $\Lambda ^{\left( 4\right) }\left( \rho \right) $ in (\ref{n4})
becomes,%
\begin{equation}
\Lambda _{\text{recover}}^{\left( 4\right) }\left( \rho \right) \equiv
\left( \mathcal{R\circ }\Lambda ^{(4)}\right) \left( \rho \right) \overset{%
\text{def}}{=}\sum_{k=0}^{15}\dsum\limits_{l=1}^{2}\left( R_{l}A_{k}^{\prime
}\right) \rho \left( R_{l}A_{k}^{\prime }\right) ^{\dagger }\text{,}
\end{equation}

\emph{Entanglement Fidelity}. We want to describe the action of $\mathcal{%
R\circ }\Lambda ^{(4)}$ restricted to the code subspace $\mathcal{C}$.
Therefore, we compute the $2\times 2$ matrix representation $\left[
R_{l}A_{k}^{\prime }\right] _{|\mathcal{C}}$ of each $R_{l}A_{k}^{\prime }$
with $l=1$, $2$ and $k=0$,.., $15$ and it turns out that the only matrices $%
\left[ R_{l}A_{k}^{\prime }\right] _{|\mathcal{C}}$ with non-vanishing trace
are given by,%
\begin{equation}
\left[ R_{1}A_{r}^{\prime }\right] _{|\mathcal{C}}=\sqrt{\tilde{p}%
_{r}^{\left( 4\right) }}\left( 
\begin{array}{cc}
1 & 0 \\ 
0 & 1%
\end{array}%
\right) \text{,}
\end{equation}%
for $r=0$, $5$, $6$, $7$, $8$, $9$, $10$, $15$. Therefore, the entanglement
fidelity $\mathcal{F}_{DFS}^{\left( 4\right) }\left( \mu \text{, }p\right) $
defined as,%
\begin{equation}
\mathcal{F}_{DFS}^{\left( 4\right) }\left( \mu \text{, }p\right) \overset{%
\text{def}}{=}\mathcal{F}_{DFS}^{\left( 4\right) }\left( \frac{1}{2}%
I_{2\times 2}\text{, }\mathcal{R\circ }\Lambda ^{(4)}\right) =\frac{1}{%
\left( 2\right) ^{2}}\sum_{k=0}^{15}\dsum\limits_{l=1}^{2}\left\vert \text{tr%
}\left( \left[ R_{l}A_{k}^{\prime }\right] _{|\mathcal{C}}\right)
\right\vert ^{2}\text{,}
\end{equation}%
is given by,%
\begin{equation}
\mathcal{F}_{DFS}^{\left( 4\right) }\left( \mu \text{, }p\right) =\tilde{p}%
_{0}^{\left( 4\right) }+\tilde{p}_{5}^{\left( 4\right) }+\tilde{p}%
_{6}^{\left( 4\right) }+\tilde{p}_{7}^{\left( 4\right) }+\tilde{p}%
_{8}^{\left( 4\right) }+\tilde{p}_{9}^{\left( 4\right) }+\tilde{p}%
_{10}^{\left( 4\right) }+\tilde{p}_{15}^{\left( 4\right) }\text{.}
\end{equation}%
After some algebra, it follows that,%
\begin{eqnarray}
\mathcal{F}_{DFS}^{\left( 4\right) }\left( \mu \text{, }p\right) &=&\mu
^{3}\left( -8p^{4}+16p^{3}-10p^{2}+2p\right) +\mu ^{2}\left(
24p^{4}-48p^{3}+28p^{2}-4p\right) +  \notag \\
&&  \notag \\
&&+\mu \left( -24p^{4}+48p^{3}-30p^{2}+6p\right) +\left(
8p^{4}-16p^{3}+12p^{2}-4p+1\right) \text{.}
\end{eqnarray}%
Finally, following the same line of reasoning presented above, $\mathcal{F}%
_{DFS}^{\left( 5\right) }\left( \mu \text{, }p\right) $ and $\mathcal{F}%
_{DFS}^{\left( 6\right) }\left( \mu \text{, }p\right) $ can be computed as
well.


\begin{thebibliography}{99}
\bibitem{nielsen00} M.\ A. Nielsen and I. L. Chuang, "\emph{Quantum
Computation and Information}", Cambridge University Press (2000).

\bibitem{laflamme07} P. Kaye, R. Laflamme and M. Mosca, "\emph{An
Introduction to Quantum Computing}", Oxford University Press (2007).

\bibitem{perimeter} D. Gottesman, "\emph{An Introduction to Quantum Error
Correction and Fault-Tolerant Quantum Computation}",
arXiv:quant-ph/0904.2557 (2009).

\bibitem{knill97} E. Knill and R. Laflamme, "\emph{Theory of quantum
error-correcting codes}", Phys. Rev. \textbf{A55}, 900 (1997).

\bibitem{calderbank97} A. R. Calderbank et \textit{al}., "\emph{Quantum
Error Correction and Orthogonal Geometry}", Phys. Rev. Lett. 78, 405 (1997).

\bibitem{zanardi97} P. Zanardi and M. Rasetti, "\emph{Error Avoiding Quantum
Codes}", Mod. Phys. Lett. \textbf{B11}, 1085 (1997).

\bibitem{zanardi97+} P. Zanardi and M. Rasetti, "\emph{Noiseless Quantum
Codes}", Phys. Rev. Lett. \textbf{79}, 3306 (1997).

\bibitem{lidar98} D. A. Lidar et \textit{al}., "\emph{Decoherence-Free
Subspaces for Quantum Computation}", Phys. Rev. Lett. \textbf{81}, 2594
(1998).

\bibitem{lidar99} D. A. Lidar et \textit{al}., "\emph{Concatenating
Decoherence-Free Subspaces with Quantum Error Correcting Codes}", Phys. Rev.
Lett. \textbf{82}, 4556 (1999).

\bibitem{bacon00} D. Bacon et \textit{al}., "\emph{Universal Fault-Tolerant
Computation on Decoherence-Free Subspaces}", Phys. Rev. Lett. \textbf{85},
1758 (2000).

\bibitem{arrigo08} A. D'Arrigo et. \textit{al}., "\emph{Memory effects in a
Markov chain dephasing channel}", Int. J. Quantum Info. \textbf{6}, 651
(2008).

\bibitem{shabani08} A. Shabani, "\emph{Correlated errors can lead to better
performance of quantum codes}", Phys. Rev. \textbf{A77}, 022323 (2008).

\bibitem{clemens04} J. P. Clemens et. \textit{al}., "\emph{Quantum error
correction against correlated noise}", Phys. Rev. \textbf{A69}, 062313
(2004).

\bibitem{schumacher96} B. Schumacher, "\emph{Sending entanglement through
noisy quantum channels}", Phys. Rev. \textbf{A54}, 2614 (1996).

\bibitem{lidar03} D. A. Lidar and K. B. Whaley, "\emph{Decoherence-Free
Subspaces and Subsystems}", arXiv:quant-ph/0301032 (2003).

\bibitem{nielsen96} M. A. Nielsen, "\emph{The entanglement fidelity and
quantum error correction}", arXiv: quant-ph/9606012 (1996).

\bibitem{knill02} E. Knill et \textit{al}., "\emph{Introduction to Quantum
Error Correction}", arXiv:quant-ph/020717 (2002).
\end{thebibliography}
\end{document}